\documentclass[preprintnumbers,superscriptaddress,amsmath,amssymb,prb,twocolumn,floatfix]{revtex4}

\usepackage{graphicx}               
\usepackage{psfrag}
\usepackage{dcolumn}                
\usepackage{bm}                     
\usepackage{subfigure}

\newcommand{\ba}{\begin{array}}
\newcommand{\ea}{\end{array}}
\newcommand{\bc}{\begin{center}}
\newcommand{\ec}{\end{center}}

\newcommand{\eps}{\epsilon}

\newcommand{\UA}{\uparrow}
\newcommand{\DA}{\downarrow}

\newcommand{\mustbe}{\stackrel{!}{=}}

\newcommand{\SC}{{\cal S}}

\begin{document}

\newcommand{\Eq}[1]{{Eq.~(\ref{#1})}}
\newcommand{\EQ}[1]{{Equation~(\ref{#1})}}
\newcommand{\av}[1]{{\left<{#1}\right>}}
\newcommand{\E}{{\textrm{e}}}
\newcommand{\note}[1]{.\newline\marginpar{\LARGE\bf!}{\bf #1}\newline}

\newcommand{\cdag}{c^\dagger}
\newcommand{\cnod}{c^{\phantom{\dagger}}}
\newcommand{\adag}{a^\dagger}
\newcommand{\anod}{a^{\phantom{\dagger}}}
\newcommand{\ctdag}{\tilde c^\dagger}
\newcommand{\ctnod}{\tilde c^{\phantom{\dagger}}}

\title{Ferromagnetic Polarons, Phase Separation, Stripes and Polaron Lattice:\\
  The two- and three-dimensional ferromagnetic Kondo Model}

\author{Maria Daghofer}
\affiliation{Institute for Theoretical and Computational Physics,
   Graz University of Technology, Petersgasse 16, A-8010 Graz,
   Austria.}
\email{daghofer@itp.tu-graz.ac.at}
\author{Winfried Koller}
\affiliation{Department of Mathematics,
   Imperial College, 180 Queen's Gate, London SW7 2BZ, UK.}
\author{Hans Gerd Evertz}
\author{Wolfgang von der Linden}
 \affiliation{Institute for Theoretical and Computational Physics,
   Graz University of Technology, Petersgasse 16, A-8010 Graz,
   Austria.}

\date{October 10, 2003}

\begin{abstract}
We investigate the two- and three-dimensional ferromagnetic Kondo
lattice model by unbiased Monte Carlo simulations. A phase
diagram for the two-dimensional model is presented, in which the
stability of magnetic order and ferromagnetic polarons is examined
with respect to the antiferromagnetic superexchange $J'$ and
temperature. The Monte Carlo simulations reveal that $J'\geq 0.02$
strengthens individual polarons while small $J'< 0.02$ favors larger
clusters and phase separation except for small doping. Lowering the
temperature stabilizes ferromagnetic polarons for realistic $J'\gtrsim
0.01$, while phase separation is only favored for very small $J'\lesssim
0.01$. Our Monte Carlo simulations show  that low temperatures can
lead to diagonal or vertical stripes depending on $J'$.  

Simulations for three-dimensional systems yield ferromagnetic polarons,
which form a `polaron lattice' at higher doping levels $0.2 \lesssim
x\lesssim 0.23$, when independent polarons do no longer fit into the
system. No tendency to phase separation is observed in three dimensions.
\end{abstract}



\maketitle

\section{Introduction}                                  \label{sec:intro}

Doped and undoped manganese oxides, such as R$_{1-x}$A$_x$MnO$_3$, where R
denotes a rare earth and A an alkaline earth, have a very complex
phase diagram~\cite{proceedings98,Nagaev:book} depending on
temperature, doping, ionic radius of the involved elements or
magnetization. They have been thoroughly investigated because 
of the CMR effect and have been found to have a rich phase diagram
including ferromagnetic (FM), antiferromagnetic (AFM), insulating,
metallic and charge ordered states. Quasi two-dimensional (2D) systems with
well separated MnO$_2$-(bi)layers exist. 

A comprehensive understanding of their properties certainly requires
treatment of the phononic degrees of freedom and of the orbital degeneracy
including Coulomb repulsion, which however would pose overwhelming difficulties,
especially in more than one dimension.

The much simpler FM Kondo lattice model~\cite{zener51} captures some of the properties
of these materials, namely the hopping of the itinerant manganese $e_g$ electrons via
the intermediate oxygen ions (double exchange) and their strong Hund's rule coupling to the
$S=3/2$ corespin formed by the localized manganese $t_{2g}$ electrons. An
additional antiferromagnetic coupling between the corespins is often included to
account for superexchange of the $t_{2g}$ electrons. Even for this model,
full quantum mechanical calculations are difficult in more than one dimension, and it has therefore been
proposed to replace the $t_{2g}$ corespin by a classical
spin.~\cite{gennes60,dagotto98:_ferrom_kondo_model_mangan,furukawa98} While
this approximation has been criticized,~\cite{EdwardsI,Nolting01,Nolting03} it has been 
shown to yield good results at low but finite
temperatures.~\cite{dagotto98:_ferrom_kondo_model_mangan} As the 
ferromagnetic coupling $J_H$ of the itinerant electrons to the
corespins is far larger than the hopping strength or the
antiferromagnetic superexchange $J'$, energy scales can be separated and
the model can thereby be considerably simplified. In the customary
$J_H\to \infty$ approach and in the second order perturbative
treatment of the virtual excitations\cite{KollerPruell2002a} for large
but finite $J_H$, it is assumed that the $e_g$ electrons are parallel
to the local corespin. Double occupancies are thus suppressed and
Coulomb repulsion is usually neglected. 

The FM Kondo lattice model with classical corespins has been extensively
investigated.~\cite{dagotto98:_ferrom_kondo_model_mangan,
yunoki98:_static_dynam_proper_ferrom_kondo,yunoki98:_phase,Yi_Hur_Yu:spinDE,dagotto01:review,furukawa98,
Motome_Furukawa_3dDE,Motome_Furukawa_Ucl,KollerPruell2002a,KollerPruell2002b,KollerPruell2002c,Aliaga_island_2d}
The influence of classical
phonons\cite{yunoki98:_phase_separ_induc_orbit_degrees,hotta00:_competition_fm_co,hotta01:_stripes_oo_manganites}
and disorder\cite{Motome_Furukawa_disorder,Motome_Furukawa_disorder_b} has been addressed. Two-orbital models 
have been treated, mainly with classical phonons but without Coulomb
repulsion.
Aliaga {\it et al.} have examined the two-dimensional one-orbital Kondo model
for $J_H \to \infty$
with an algorithm similar to ours\cite{Aliaga_island_2d} and reported several
phases (stripes, island phases for commensurate fillings, and the so
called ``Flux Phase'') as well as phase separation (PS), which has
also been related by other
authors.~\cite{dagotto98:_ferrom_kondo_model_mangan, 
yunoki98:_phase_separ_induc_orbit_degrees,
yunoki98:_phase,dagotto01:review,moreo_science_99}
Stripes were reported in a Kondo model applied to cuprates with
different parameter
values.\cite{Moraghebi_01:kondoCu,Moraghebi_02:kondoCu,Moraghebi_02b:kondoCu}  

In previous
papers,\cite{KollerPruell2002a,KollerPruell2002b,KollerPruell2002c} we have
studied the one-dimensional model 
by Monte Carlo techniques similar to those used before.
Careful analysis of the data showed, however,
that instead of phase separation there are 
independent small FM polarons.
We have recently extended these studies to the two-dimensional 
case\cite{DaghoferKoller2003} for doping $0<x\lesssim 12\%$ and at AFM corespin-coupling $J'=0.02$.
A thorough examination of a variety of observables
showed that, like in 1D, there is no phase separation
but holes enter the system by forming independent FM polarons with a single hole inside.

In the present paper we provide a 
phase diagram of the two-dimensional model as a function of 
corespin coupling $J'$
and hole doping up to $x=0.6$.
We thoroughly examine the transition from the polaronic 
to the homogeneous ferromagnetic phase.
At moderate to large $J'$ the polarons persist,
whereas at very small $J'$, phase separation occurs.
In the homogeneous phase, strong FM order is only present
at small to moderate $J'$.

At least for small doping, polarons are favored over phase separation by entropy, and their dependence on
temperature is therefore of interest.
Previous studies have mostly employed an inverse temperature
of $\beta=50$ (corresponding to an experimentally relevant
temperature of 50-100K, depending on hopping strength).
We show that further lowering the temperature
actually {\em strengthens} the polarons,
even though the effect of entropy is reduced, i.e., they are also
energetically favored. Only at small $J'$, low temperatures enhance
a tendency to phase separation and toward diagonal chains of holes.

We also investigate the 3D model where we likewise 
find independent polarons at small doping, 
and in addition a 'polaron lattice' at higher doping.
No phase separation is observed at any doping.

The outline of this paper is as follows.
In Sec.~\ref{sec:model} we introduce the Hamiltonian and the numerical method.
In Sec.~\ref{sec:FM_Pol_2d_3d} we establish FM model polarons in two and
three dimensions; we give an analytic estimate of their energy
compared to phase separation in Sec.~\ref{sec:muc_ps_pol} and find polarons
to dominate except for very small $J'$. Section~\ref{sec:Numerical_Results_2d}  
contains unbiased MC results for the two dimensional model.
We present a phase diagram for $\beta=50$ in Sec.~\ref{sec:phase_diagram}.

We study temperature effects in Sec.~\ref{MC:beta}, and a phase diagram for lower temperature $\beta=80$
($30K - 60K$) is presented in Sec.~\ref{MC:beta:phase_diagram}.

Finally, we present unbiased MC data on the three dimensional model at
$0<x\lesssim 0.4$ in Sec.~\ref{sec:Numerical_Results_3d}.

\section{Model Hamiltonian and Method}        \label{sec:model}

In this paper, we treat the ferromagnetic Kondo lattice model in two
and three dimensions with one orbital and classical corespins.
We treat configurations with the electron spin antiparallel to the
local corespin in second order perturbation theory, as proposed in
Ref.~\onlinecite{KollerPruell2002a}, which is a systematic improvement
over the $J_H \to \infty$ approach. The resulting effective spinless 
fermion (ESF) Hamiltonian is
\begin{equation}                                       \label{eq:H}
  \hat H = -\sum_{<i,j>} t^{\UA\UA}_{i,j}\,
    \cdag_{i}\,\cnod_{j} - \sum_{i,j}
    \frac{t^{\UA\DA}_{i,j}\,t^{\DA\UA}_{j,i}}{2J_\textrm{H}}\, \cdag_{i}\cnod_{i}
    + J'\sum_{<i,j>} \mathbf S_i \cdot \mathbf S_j \;.
\end{equation}
The spinless fermion operator $\cnod_{j}$ ($\cdag_{i}$) destroys (creates) a
\emph{local} spin-up electron (i.e. an electron with spin parallel to
the corespin at the same site) at site $i$. Down electrons
have been integrated out, and the spin index has therefore been
omitted. 
Since the local corespins may point in an arbitrary direction,
the ESF model~(\ref{eq:H}) still contains contributions from both 
spin orientations with respect to a {\em global} spin-quantization axis.
$J_H$ is the large ferromagnetic coupling of the $e_g$ electron spin
to the corespin and is set to $J_H=6$ throughout this work. The small AFM
superexchange parameter $J'$ takes values between $J'=0$ and
$J'=0.05$ for manganites. 

The hopping strength $t^{\UA\UA}_{i,j}$ in the kinetic energy (first term of
Eq.\ref{eq:H}) depends on the relative angle $\vartheta_{i,j}$
between the corespins at the two neighboring sites:
\begin{multline}                                      \label{eq:modihop}
  t^{\UA\UA}_{i,j}\;=\;
  t_{0}\; u_{i,j}^{\UA\UA}\; =\\= t_0(c_ic_j + s_is_j\E^{i(\phi_j - \phi_i)})=
    t_0\cos(\vartheta_{ij}/2)\;\E^{i\psi_{ij}},
\end{multline}
where $c_i=\cos(\vartheta_i/2)$ and $s_i=\sin(\vartheta_i/2)$. The factor depends on the
polar coordinates $\vartheta$ and $\phi$ of the corespins and contains a
complex phase factor, which can lead among others to the so called ``Flux Phase'', see
Refs.~\onlinecite{Aliaga_island_2d,Agterberg_00,Yamanaka_98} and
Sec.~\ref{sec:fm_pm_flux}. We use the parameter $t_0=1$ as unit of energy
throughout this paper.

The second term in Eq.~\ref{eq:H} contains the second order treatment of the
virtual excitations, where an up-electron at site $i$ can become a
down-electron at site $j$ and immediately hop back to become an
up-electron at site $i$ again. Its strength also depends on the corespins but is
always real:
\begin{equation}                                      \label{eq:modisup}
  \begin{aligned}
  t^{\UA\DA}_{i,j}\;t^{\DA\UA}_{j,i}& = t_{0}^2\ |u_{i,j}^{\DA\UA}|^2 =\\
  & =t_0( |c_is_j\E^{-i\phi_j} + s_ic_j\E^{-i\phi_i}|^2) = t_0\sin(\vartheta_{ij}/2)^2\,.
  \end{aligned}
\end{equation}
We treat this Hamiltonian by unbiased Monte Carlo simulations for the grand
canonical ensemble using the standard
algorithm~\cite{dagotto98:_ferrom_kondo_model_mangan} and for the canonical
ensemble with the algorithm proposed in
Ref.~\onlinecite{DaghoferKoller2003}. 
The lattice size is $12\times 14$ sites in 2D and $6\times 6\times 4$ in 3D
where not otherwise indicated. We mainly choose non-quadratic lattices
in order to minimize finite size effects due to closed shells.
For our MC results, 50 to 200 lattice sweeps have been skipped between
measurements and autocorrelations have been analyzed. Wherever they
are found to be very long (notably around the phase boundary between the
Flux phase and the FM/PM phase and for the vertical stripe phase),
parallel tempering\cite{huknem:95,huknem:96,mar:96} has been employed. 

\section{FM Model-Polarons in 2D and 3D}               \label{sec:FM_Pol_2d_3d}
In Ref.~\onlinecite{DaghoferKoller2003}, we showed that the results of unbiased
Monte Carlo simulations for $J' = 0.02,\ \beta=50$ and $J_\textnormal{H}=6$
for doping levels up to $x\approx 12\%$ (i.e. near the completely
filled lower Kondo band) correspond
to independent ferromagnetic polarons and not to phase separation.
To this end, we constructed a model of independent polarons, evaluated several
observables (spectral density, DOS, corespin correlation and dressed corespin correlation
\Eq{eq:hss_def}) for this model, compared the results to the MC data and
found almost perfect agreement.

\begin{figure}
  \centering
  \subfigure[]{\includegraphics[width=0.23\textwidth]{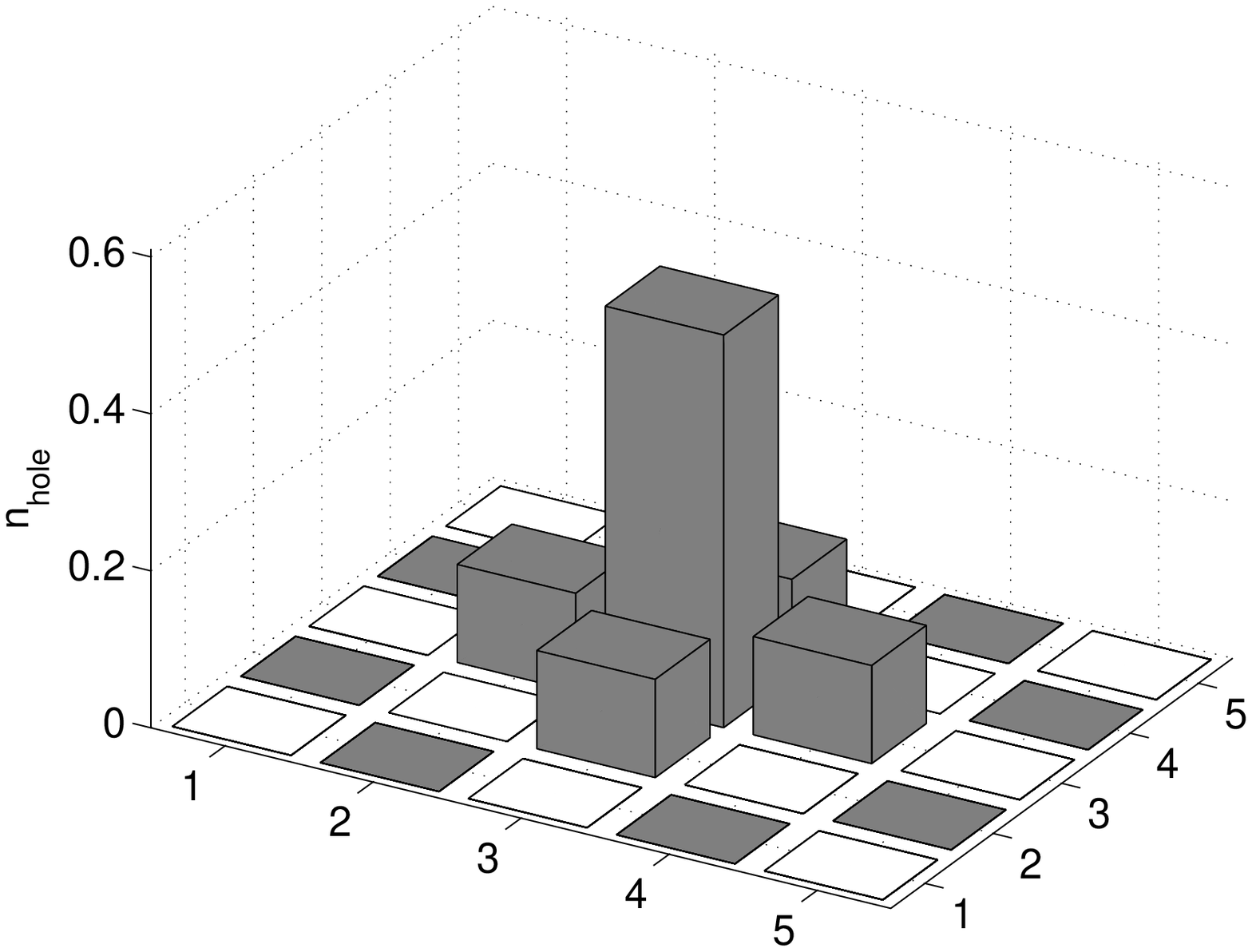}\label{polaron_conf}}\hfill
  \subfigure[]{\includegraphics[width=0.23\textwidth]{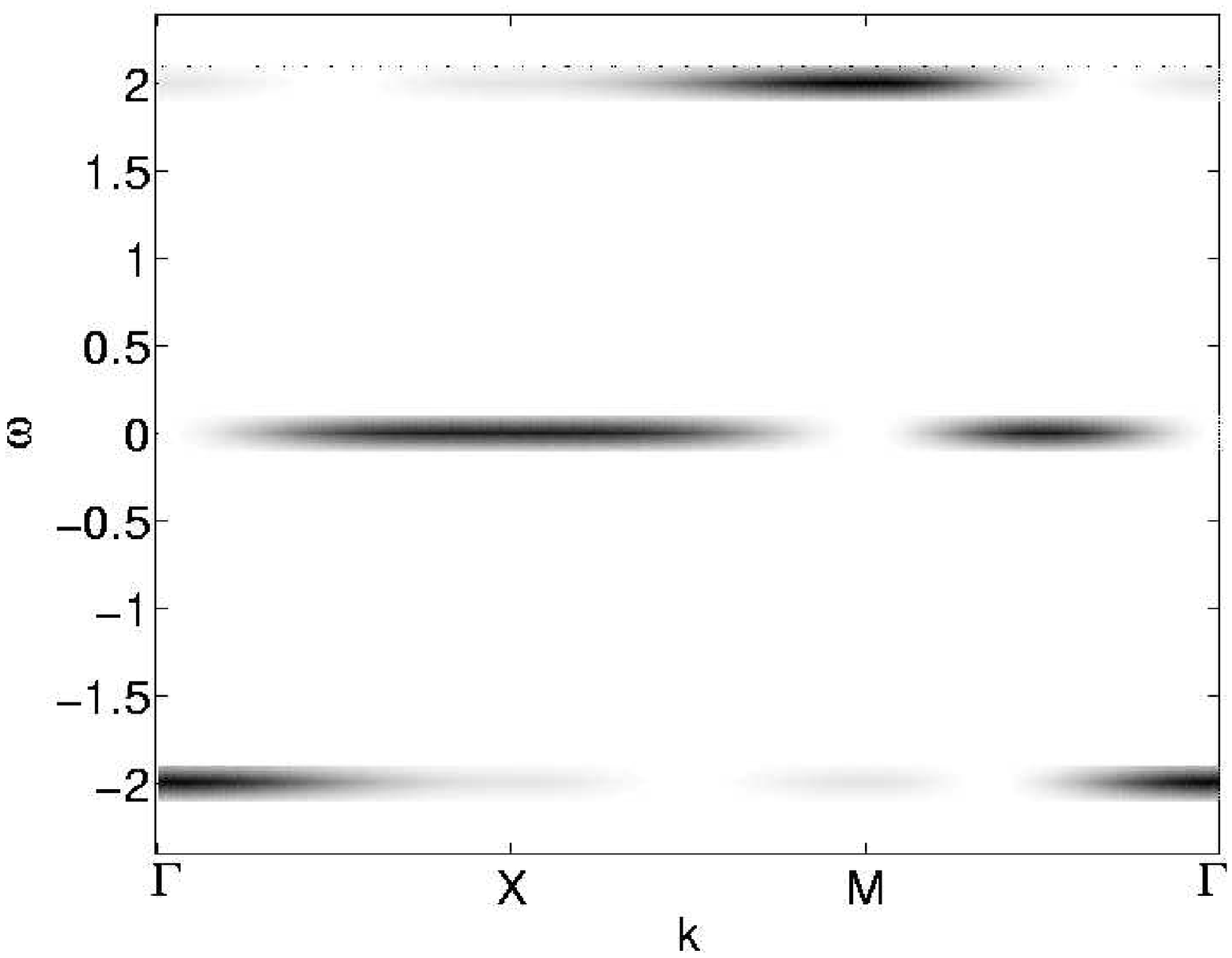}\label{polaron_bands}}\\
  \caption{Idealized two-dimensional FM polaron of $L_\textrm{f}=5$ lattice sites,
  embedded in an AFM background (shown in Ref.~\onlinecite{DaghoferKoller2003})
  (a) Spin and hole-density configuration for the groundstate. Empty
  (filled) squares represent spin down (up). Height represents hole
  density.
  (b) Contribution of the polaron to the one-particle spectral
  function. For visibility, the $\delta$-peaks in the spectral density
  have been broadened to a width of $0.2$.}
  \label{polaron}
\end{figure}

In the undoped case $x=0$, the corespins are aligned
antiferromagnetically because of virtual excitations into the upper
Kondo band and because of the AFM superexchange $J'$. The effective
strength of the antiferromagnetic interaction is given by
$J_\textrm{eff} = J'+\tfrac{1}{2J_H}$ for $n\approx 1$.\cite{KollerPruell2002a}
One FM polaron in 2D consists of just one spin flipped from the AFM
background and thus forming a five-site FM region where a hole is trapped,
see~Fig.~\ref{polaron_conf}. The eigenstates of the hole inside this small
FM cluster lead to signals at $\omega=\pm 2$ and at $\omega=0$ in the spectral density and the
DOS, see Fig.~\ref{polaron_bands}.
For a large number of polarons, i.e., of flipped spins, the FM
polarons occasionally overlap and thereby form larger FM areas; such
overlaps are neither suppressed nor encouraged in the
independent polaron model.

\begin{figure}
  \centering
   \begin{minipage}{0.1\textwidth}
     $1^\textrm{st}$ layer\\[4em]
     $2^\textrm{nd}$ layer\\[4em]
     $3^\textrm{rd}$ layer
   \end{minipage}
   \begin{minipage}{0.3\textwidth}
     \includegraphics[width=0.55\textwidth]{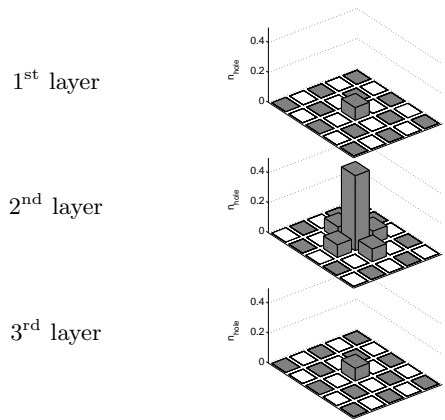}
   \end{minipage}
  \caption{Spin and hole-density configuration of an
    idealized three-dimensional FM polaron of $L_\textrm{f}=7$ lattice sites,
  embedded in an AFM background. The three layers represent three consecutive
  plains in the 3D lattice. Empty
  (filled) squares represent spin down (up). Height represents hole
  density in the ground state.\label{polaron_3d}}
\end{figure}

In 3D, the situation is very similar: When one spin is flipped in the AFM
lattice, a FM domain of seven sites results. This is schematically
depicted in Fig.~\ref{polaron_3d}. Its eigenenergies are given by
$\omega = 0 $ (fourfold degenerate) and $\omega = \pm \sqrt{6}$. When
a hole is inserted into the 3D lattice, it can gain the energy
$-\sqrt{6}$, but the energy $12 J_\textrm{eff}$ for the breaking of six
AFM bonds has to be paid. The polaron energy is therefore given by
\begin{equation}                                       \label{eq:e_pol}
  e_\textrm{pol} = -\sqrt{6} + 12\ J_\textrm{eff}\;.
\end{equation}

This energy determines the critical chemical potential
$\mu^*=-e_\textrm{pol}$, at which holes that form polarons enter the
completely filled lower Kondo band of a 3D system. Interestingly, such
seven-site stars were also reported to result from combined lattice and spin
effects for the electron doped system, i.e., for few electrons.\cite{allen01:sl_polaron}

\subsection{Energy comparison of Phase Separation and FM Polarons} \label{sec:muc_ps_pol}

We examine the competition of phase separation and FM polarons by comparing
their energies for the relevant range of the effective antiferromagnetic
superexchange $J_\textrm{eff}=J'+\tfrac{1}{2J_H}$, which can of course only be done
approximately.
We obtain the critical chemical potential $\mu_\textrm{PS}$ for phase separation by setting
the energy of the completely filled AFM band equal to that of a partially
filled FM band.

\begin{equation}\label{eq:muc_ps}
-\mu_\textrm{PS}-z J_\textrm{eff} \mustbe \underbrace{\int_{-z}^{\mu_\textrm{PS}}\!\!\!\!\epsilon\;
 n(\epsilon)\; d\epsilon}_{e_\textrm{kin}} - \mu_\textrm{PS}\int_{-z}^{\mu_\textrm{PS}}\!\!\!\!n(\epsilon)\; d\epsilon\;,
\end{equation}
where $z$ is the number  of nearest neighbors for each site (4 in 2D,
6 in 3D) and $n(\epsilon)$ is the one-particle density of states of a
tight binding band with $\epsilon(k_x,k_y)=-2\cos(k_x)-2\cos(k_y)$ or
$\epsilon(k_x,k_y,k_z)=-2\cos(k_x)-2\cos(k_y)-2\cos(k_z)$, respectively.
The kinetic energy per site $e_\textrm{kin}$ (first term on the
right-hand side) is zero for the completely filled band. 

\begin{figure}
\subfigure{\includegraphics[width=0.235\textwidth]{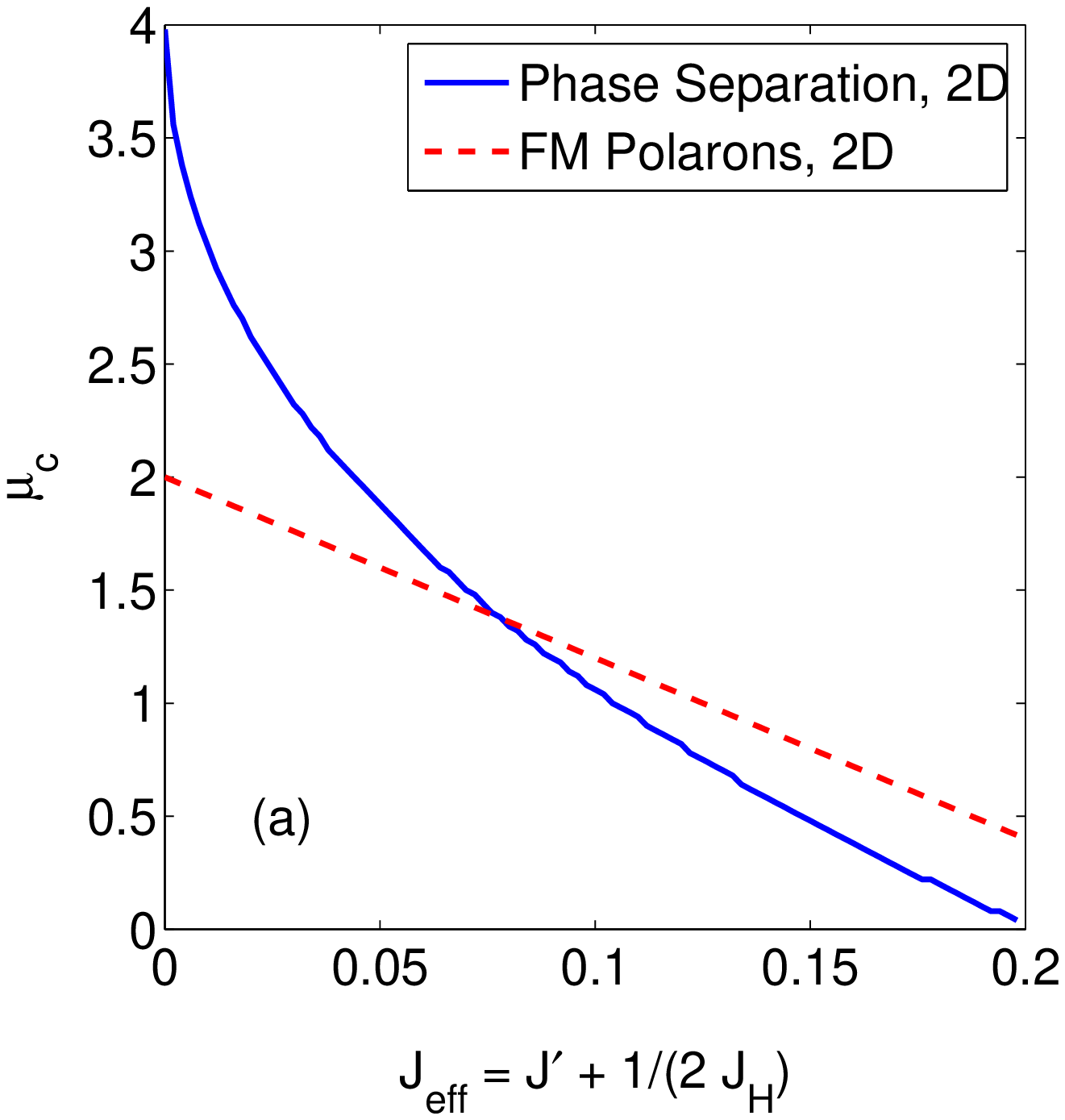}\label{fig:muc_PS_Pol_2d}}
\subfigure{\includegraphics[width=0.235\textwidth]{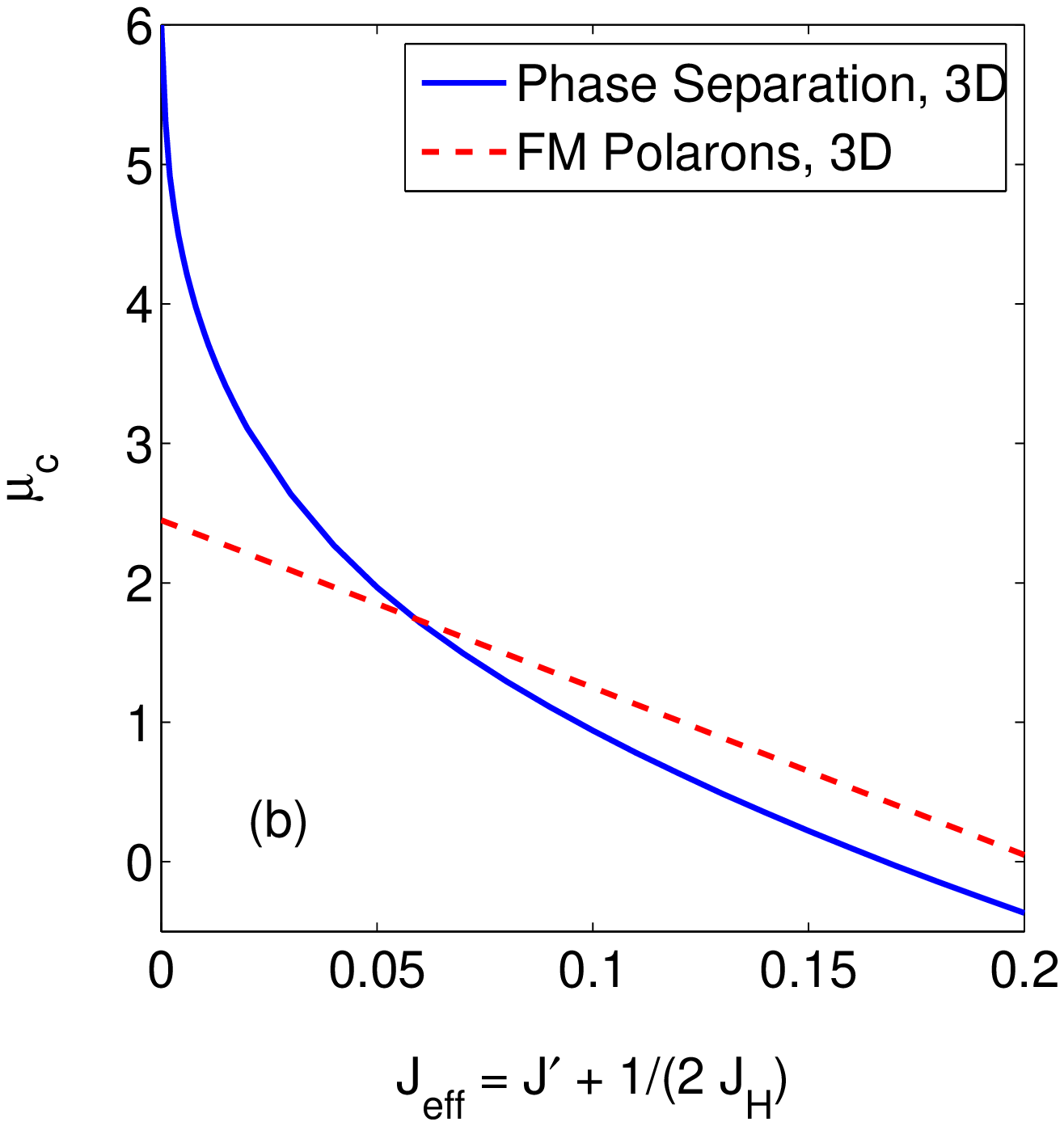}\label{fig:muc_PS_Pol_3d}}
\caption{(Color online) Critical chemical potential for phase separation (solid line) and FM
  polarons of the shape depicted in Fig.~\ref{polaron_conf} for (a): 2
  dimension and (b) 3 dimensions.\label{fig:muc_PS_Pol}}
\end{figure}

The critical chemical potential for FM polarons is given by
$\mu_\textrm{pol} = 2-8 J_\textrm{eff}$ in 2D (see
Ref.~\onlinecite{DaghoferKoller2003}) and $\mu_\textrm{pol} = \sqrt{6}-12 J_\textrm{eff}$
in 3D (see Eq.\ref{eq:e_pol}). It
is compared to $\mu_\textrm{PS}$ in Fig.~\ref{fig:muc_PS_Pol} for two and
three dimensions. In contrast
to the energy minimization performed for 1D systems (given in Ref.~\onlinecite{KollerPruell2002c})
we did not optimize the shape of the polarons, but we always used polarons
consisting of one single flipped spin as depicted in Figs.~\ref{polaron_conf}
and~\ref{polaron_3d}. This puts FM polarons slightly at a disadvantage.

For $\mu$ larger than the critical values, the band is completely filled and
AFM. When the chemical potential is lowered and holes are introduced into the
system, this leads to PS if $\mu_\textrm{PS}>\mu_\textrm{pol}$ and to
polarons if $\mu_\textrm{pol} > \mu_\textrm{PS}$.

For 2D, one sees that PS is favored for small
$J_\textrm{eff}\lesssim 0.08$ in accordance with the MC data, see
Sec.~\ref{MC:J_0}. For the parameter $J_H=6$ used in our calculations, $J_\textrm{eff}\approx0.083$
corresponds to $J'=0$, where we do indeed observe phase separation. For
larger values of $J_\textrm{eff}$, i.e., smaller $J_H < 6$ or larger $J' > 0$,
polarons are not only favored by entropy, but also by energy.

In three dimensions, the preference for polarons over PS sets in for even
smaller values for $J_\textrm{eff}$.

\section{MC Results for 2D}                 \label{sec:Numerical_Results_2d}
We first investigate the influence of AFM superexchange $J'$ and doping $x$
at $\beta=50$, eventually leading to the phase diagram in
Fig.~\ref{fig:phase_diagram}. Temperature effects are discussed in
Sec. \ref{MC:beta}, with a phase diagram for $\beta=80$ in Fig.~\ref{fig:phase_diagram_b80}.

\subsection{$J' = 0.02$: From independent polarons to phase separation}          \label{sec:MC_J02}

In Ref.~\onlinecite{DaghoferKoller2003}, we found a very close
correspondence of the MC data with the results from the polaron model
for all observables up to doping levels of $x\approx 12\%$, where the
polarons already cover approximately $60\%$ of the lattice.
The polarons appear to be independent, with overlapping
polarons occurring at a similar rate as for the independent polaron model.

\begin{figure}
  \centering
  \includegraphics[width = 0.45\textwidth]{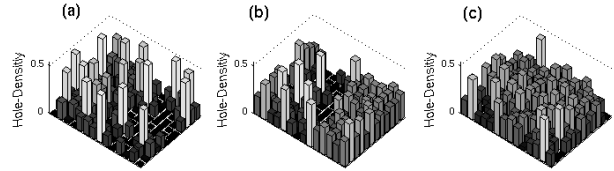}\\
  \caption{MC snapshot of the hole density for (a) $20$ holes ($x\approx
  12\%$, shown in Ref.~\onlinecite{DaghoferKoller2003}), (b) $24$ ($x\approx
    14\%$), and (c) $31$ ($x\approx 18\%$) holes in 
    a $14\times 12$  lattice at $\beta=50, J'=0.02,\; 
    J_H=6$\label{MC_snapshot_J0.02_beta80_144.eps}.
    Height represents hole density, grayshades are for better visibility.}
\end{figure}

In this section, we investigate higher doping levels $x > 12\%$ at
$\beta=50$, where the results begin to deviate from the independent polaron
results; the homogeneous FM phase sets in at $x\approx 21\%$.
In between, the polarons attract each other which leads to  phase
separation (PS). The transition from polarons to PS is not well
defined.
The polarons first coexist with large clusters, which finally
dominate. This development is illustrated in the MC snapshots depicted in
Fig.~\ref{MC_snapshot_J0.02_beta80_144.eps}. It shows snapshots for 20 holes
($x\approx 12\%$), where 19 polarons can be seen and only one hole is
delocalized, for 24 holes ($x\approx 14\%$), where polarons coexist with a
larger and more homogeneous area, and finally for 31 holes ($x\approx
18.5\%$). This last doping is only approximately 4 holes away form the
homogeneous FM phase at $x\gtrsim 21\%$, and even there, some polarons persist.

In order to check whether the addition of holes leads primarily to more small
polarons or to a growth of the existing ones, we compute a
dressed corespin correlation function
\begin{equation} \label{eq:hss_def}
  S_h(\vec r) = \frac{1}{L} \sum_{\vec i} n^h_{\vec i} S_{\vec i}\ S_{\vec i + \vec r}\;.
\end{equation}
The hole density at site $\vec i$, denoted by $n_{\vec i}^h$, is related to the
electron density via $n_{\vec i}^h=1-n_{\vec i}=1-\langle\cdag_{\vec
  i}\cnod_{\vec i}\rangle_\SC$, where $\langle n_{\vec i} \rangle_\SC$ is
the expectation value of the density given the corespin configuration $\SC$. The sum over $\vec i$ is taken over all
lattice sites and the observable is averaged over all corespin configurations
$\SC$ occurring in the MC run. The dressed correlation measures the ferromagnetic regions around
holes.

\begin{figure}
  \includegraphics[width=0.4\textwidth]{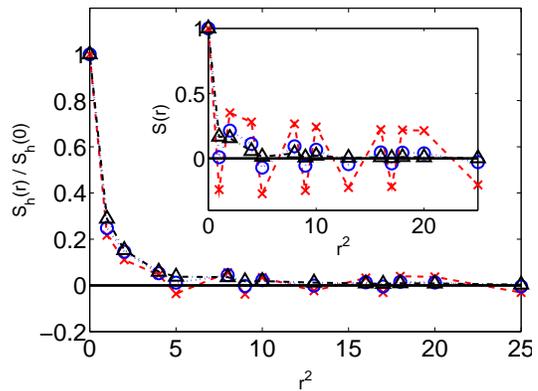}
  \caption{(Color online) Dressed corespin correlation \Eq{eq:hss_def} for $24$ holes
    ($\times$, dashed),
    $31$ holes ($\circ$, dotted) and $\approx 36$ ($\triangle$, dash-dotted)
    holes. The inset shows the usual corespin correlation in real
    space $S(\vec r) = \tfrac{1}{L} \sum_{\vec i} S_{\vec i}\ S_{\vec i + \vec r}$.
    Remaining parameters as Fig.~\ref{MC_snapshot_J0.02_beta80_144.eps}.\label{hss_ss_MC_SE0.02_pol_ps}}
\end{figure}

Figure~\ref{hss_ss_MC_SE0.02_pol_ps} shows $S_h(\vec r)$ for $24$, $31$ and
$\approx 36$ holes. The latter doping marks the point, where the more
homogeneous phase sets in and where the compressibility becomes much smaller
than in the polaronic and phase separated regimes, so that this filling can
be obtained in the grand canonical ensemble. One sees that the ferromagnetic
regions around the holes grow slightly with doping, e.~g., the correlation at $r^2=5$
goes from AFM to FM. This is an indication for phase
separation. The usual corespin correlation in real space is shown in the
inset. It shows the evolution of short-range ferromagnetic
correlations. However, no long-range FM correlations are present even for
$\approx 36$ holes ($x\approx21\%$). This shows that the orientation of the
small FM clusters is largely independent, see also the spin structure factor
in Sec.~\ref{sec:fm_pm_flux}. We therefore do not have coexistence of
macroscopic FM and AFM phases at any point for $J'=0.02$.

\subsection{Dependence of Polarons and Phase Separation on $J'$}    \label{MC:J_0}

Although $J'$ is usually taken to be very small compared to the other parameters
in the Hamiltonian, it has a considerable influence on the polaronic
regime, because it stabilizes the AFM background.

\begin{figure}[htbp]
  \centering
  \includegraphics[width = 0.45\textwidth]{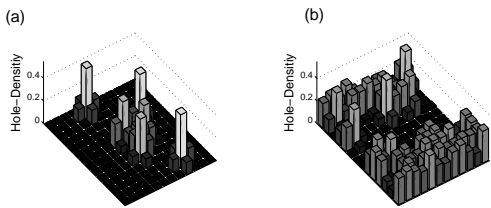}\\
  \caption{Monte Carlo snapshots for $J'=0$ and $\beta=50$. (a) $6$
  holes, polarons; (b) 20 holes, phase separation.\label{fig:MC_snapshots_J0.0}
      The MC simulations were done for $J_\textnormal{H}=6$, $\beta=50$ on a
     $12 \times 14$ lattice. Height represents hole density, grayshades are for
    better visibility.}
\end{figure}

For $J' = 0$ and $\beta = 50$, MC snapshots shown in
Fig.~\ref{fig:MC_snapshots_J0.0} reveal polarons with a
tendency to form clusters for a few holes and phase separation for larger
doping.

\begin{figure}[htbp]
  \centering
  \includegraphics[width = 0.4\textwidth]{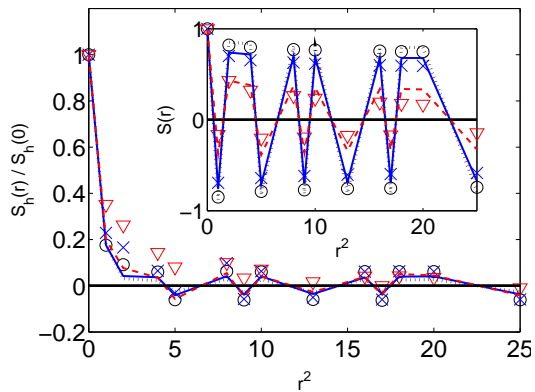}
   \caption{(Color online) Dressed corespin correlation \Eq{eq:hss_def} from unbiased MC
     Data for $J'=0$ (symbols) and $J'=0.02$ (continuous lines) for 1
    ($\circ$, dotted), 6 ($\times$, solid) and 20 ($\triangledown$, dashed)
    holes.
    The inset shows the corespin correlation. Parameters as in Fig.~\ref{fig:MC_snapshots_J0.0}.
  \label{fig:hss_MC00vsMC02}}
\end{figure}

The dressed spin correlation \Eq{eq:hss_def} for 1, 6 and 20 holes is depicted in
Fig.~\ref{fig:hss_MC00vsMC02} for $J'=0$ and $J'=0.02$. In this doping range,
the ferromagnetic regions around the holes
do not grow with doping for $J'=0.02$ (continuous lines), instead, the nearest AFM
correlation at $r^2=5$ remains AFM, which indicates polarons.
In contrast, the FM regions do grow for $J'=0$ (symbols), this growth suggests larger FM regions,
i.e. phase separation.
Also, the antiferromagnetic correlations shown in the inset decrease
faster for $J'=0$ than for $J'=0.02$.

\begin{figure}
  \subfigure{\includegraphics[width=0.235\textwidth]
    {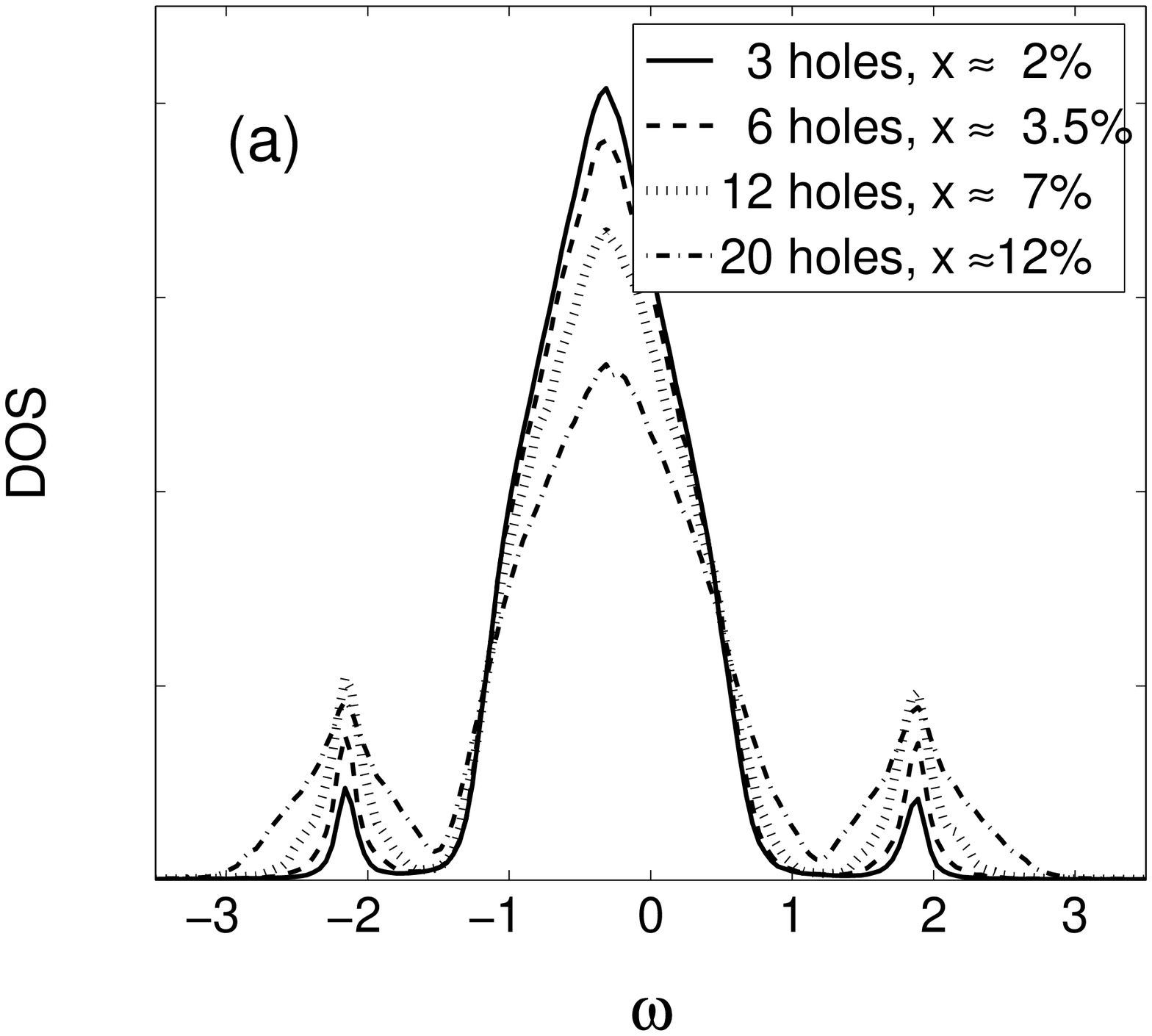}\label{dos_beta50_Nvar_02}}
  \subfigure{\includegraphics[width=0.235\textwidth]
    {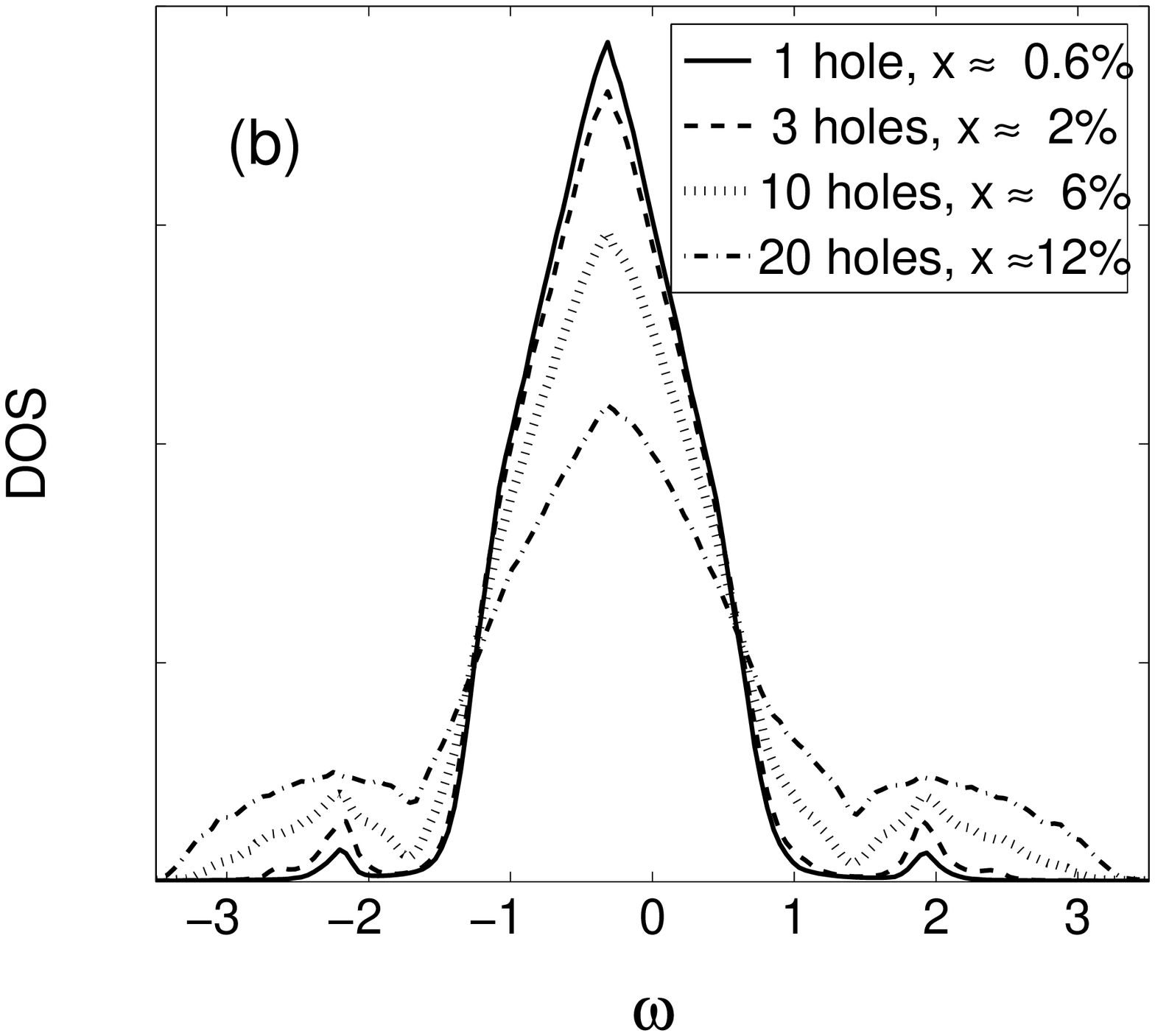}\label{dos_beta50_Nvar_0}}
  \caption{One-particle DOS for (a) $J'=0.02$ (shown in
  Ref. \onlinecite{DaghoferKoller2003}) and for (b) $J'=0$, 
  other parameters as in Fig.~\ref{fig:MC_snapshots_J0.0}.\label{dos_beta50_Nvar.eps}}
\end{figure}

Fig.~\ref{dos_beta50_Nvar.eps} shows the DOS for $J'=0.02$ and for $J'=0$ at various fillings.
For both values of $J'$ and all depicted doping ranges, one sees the
central band which stems from the movement of the electrons in the
imperfect AFM background. This band is completely filled. 
For small doping, one sees the polaronic states at
$\omega \approx \pm 2$, see also Fig.~\ref{polaron}. For $J'=0.02$,
the weight of these peaks increases with doping, but 
their shape changes only for very large doping, in accordance with
independent polaron results.\cite{DaghoferKoller2003} The pseudogap remains
clearly visible. For $J'=0$, one also sees the polaronic states and the
pseudogap for small doping, but the pseudogap is filled at larger
doping. The states filling the pseudogap stem from larger FM clusters,
and these are larger and more frequent than can be accounted for by
occasionally overlapping independent polarons.
The actual experimental observation of the
pseudogap\cite{dessauI,dessauII,dessauIII,Park95} 
rather favors the polaron scenario as opposed to PS.

\begin{figure}[htbp]
  \begin{minipage}{0.07\textwidth}
    (a)\\[2em]
    J'=0\\[9em]
    (b)\\[2em]
    J'=0.02\\[9em]
    (c)\\[2em]
    J'=0.05
  \end{minipage}
  \begin{minipage}{0.38\textwidth}
  \centering
  \subfigure{\includegraphics[width = 0.99\textwidth,trim= 0 50 0 50,clip]
    {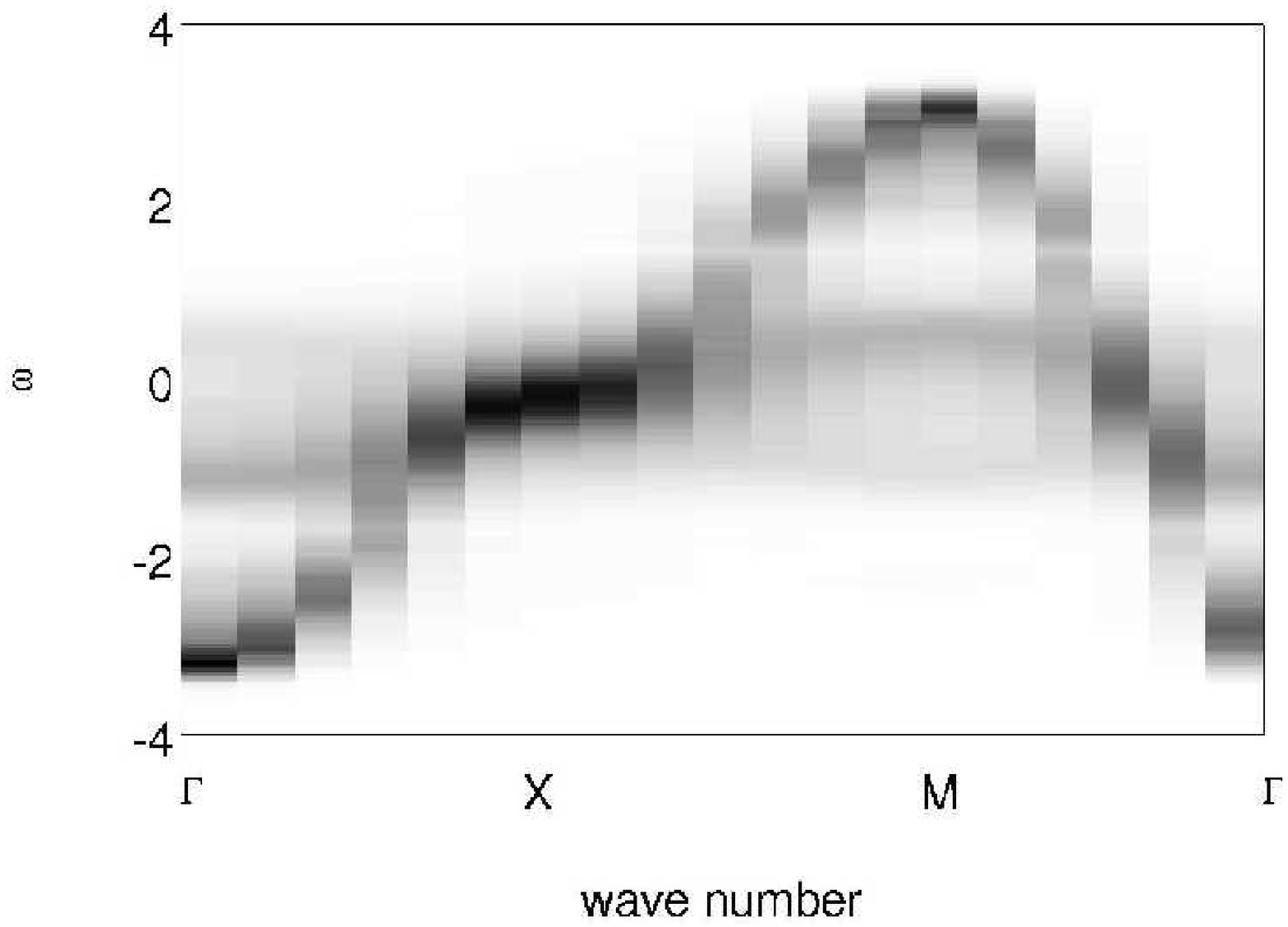}\label{fig:spec_beta50_N148_0}}
  \subfigure{\includegraphics[width = 0.99\textwidth,trim= 0 50 0 50,clip]
    {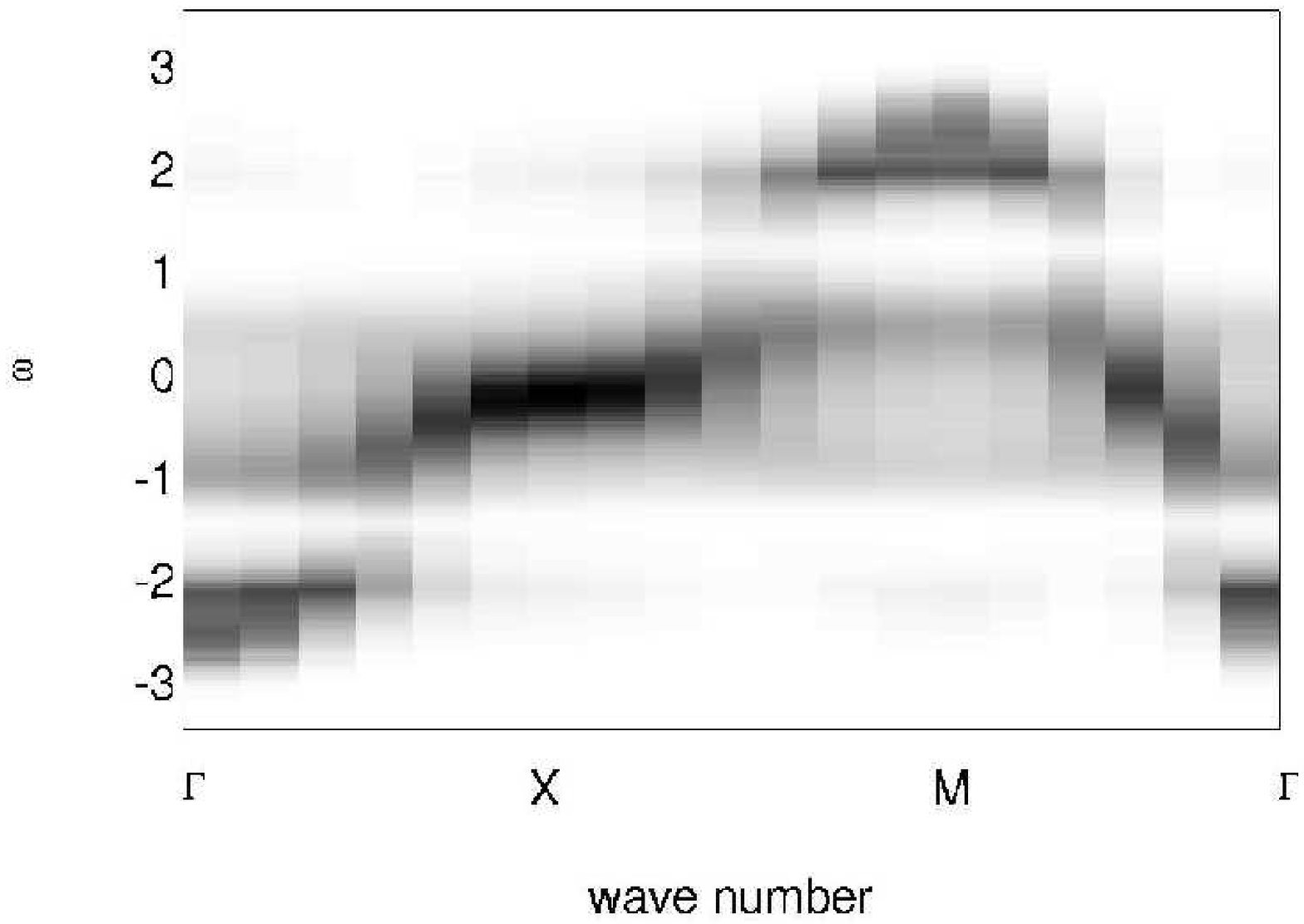}\label{fig:spec_beta50_N148_02}}
  \subfigure{\includegraphics[width = 0.99\textwidth,trim= 0 50 0 50,clip]
    {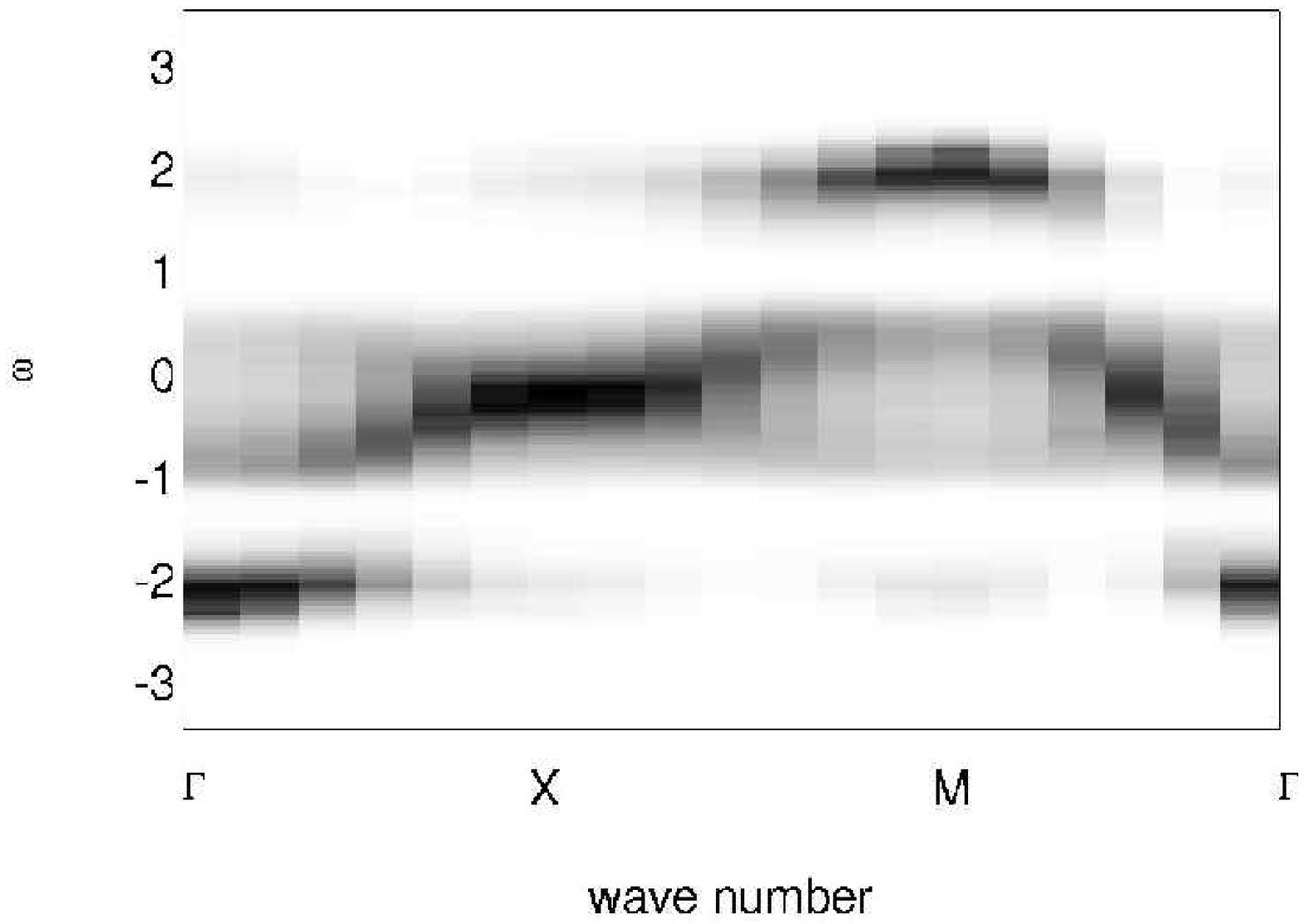}\label{fig:spec_beta50_N148_05}}
  \end{minipage}
  \caption{Spectral density for $x\approx 12\%$ and (a) $J'=0$ ($20$ holes),
  (b) $J'=0.02$ ($20$ holes, also shown in
  Ref.~\onlinecite{DaghoferKoller2003}), and (c) $J'=0.05$
  ($21$ holes) on a 
  $12 \times 14$ lattice, $\beta=50$, $J_\textnormal{H}=6$.
  \label{fig:spec_beta50_N148}}
\end{figure}

Figure~\ref{fig:spec_beta50_N148} shows the spectral density for $J'=0$,
$J'=0.02$, and $J'=0.05$ for $20$ or $21$ holes, i.e. $x\approx12\%$.
The larger FM areas for $J'=0$ [top panel (a)] lead to a second band in
addition to the one stemming from the AFM parts of the lattice. It has
a larger bandwidth, because the corespin correlations are
ferromagnetic and the hopping strength is therefore larger. The
central panel (b) shows the results for the
independent polarons at $J'=0.02$ (also given in
Ref.~\onlinecite{DaghoferKoller2003}). Around the polaronic states at
$\omega=\pm2$, one sees the signals from overlapping polarons, which
are also there for the independent polaron
model.\cite{DaghoferKoller2003} The signals from the small FM clusters
do not form a continuous band, but are instead separated from the AFM
band by the pseudogap.

A stronger antiferromagnetic superexchange $J' = 0.05$
suppresses configurations with overlapping polarons, as can be seen in
the lowest panel, Fig.~\ref{fig:spec_beta50_N148_05}. In addition to
the antiferromagnet, the data for $J' = 0.05$ show almost only one-polaron
states and the pseudogap is more pronounced. This shows that not even overlapping polarons exist for $J' =
0.05$, let alone phase separation. The reason for this behavior is that
larger $J'$ favors antiferromagnetically stacked polarons over larger
clusters. This can be seen in another dressed corespin correlation, which
takes into account hole densities at both lattice sites:

\begin{equation} \label{eq:hhss_def}
  S_{hh}(\vec r) = \frac{1}{L} \sum_{\vec i} n^h_{\vec i} S_{\vec i}\ 
  n^h_{\vec i + \vec r} S_{\vec i + \vec r}\;.
\end{equation}

\begin{figure}
  \includegraphics[width = 0.4\textwidth]{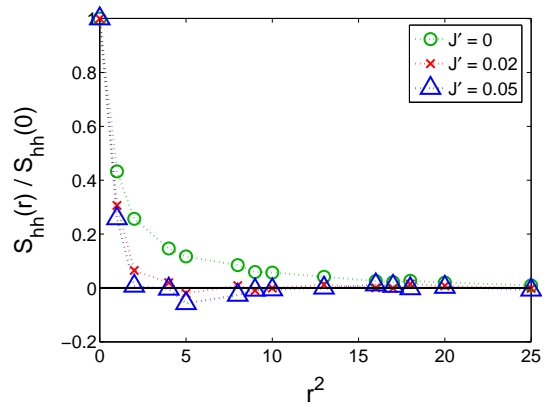}
  \caption{(Color online) Hole-spin-hole-spin correlation as defined in
    Eq.\ref{eq:hhss_def} for 20 holes  and various values of $J'$:
    Other parameters as in Fig.~\ref{fig:MC_snapshots_J0.0}}
  \label{fig:hhss_Jsevar_N148}
\end{figure}

Note that even without the corespins, this would
\emph{not} give the usual density correlation,
because it measures $n_{\vec i}n_{\vec j}=\langle\cdag_{\vec i}\cnod_{\vec i}\rangle_\SC
\langle\cdag_{\vec j}\cnod_{\vec j}\rangle_\SC$ instead of $\langle\cdag_{\vec
  i}\cnod_{\vec i}\ \cdag_{\vec j}\cnod_{\vec
  j}\rangle_\SC$. Eventually, this
observable is of course averaged over all spin configurations $\SC$ obtained in the
Monte Carlo run.
This correlation is depicted in
Fig.~\ref{fig:hhss_Jsevar_N148} for 20 holes ($x\approx12\%$) and for $J'=0,
J'=0.02$ and $J'=0.05$. One sees a positive correlation indicating PS for
$J'=0$, almost only short-range correlations over two sites (one polaron) for
$J'=0.02$ and a negative correlation at $r^2=5$ for $J'=0.05$, which
indicates antiferromagnetically stacked neighboring polarons.

\subsection{FM, PM and Flux Phase}    \label{sec:fm_pm_flux}

The exchange interaction also plays an important role at electron densities
below the polaronic and phase separated regimes, i.~e., at larger doping
levels $x \gtrsim 0.2$, which corresponds to chemical potentials
$\mu < \mu^* = -\eps_\textnormal{pol}$.

\begin{figure}
  \centering
  \includegraphics[width=0.4\textwidth]{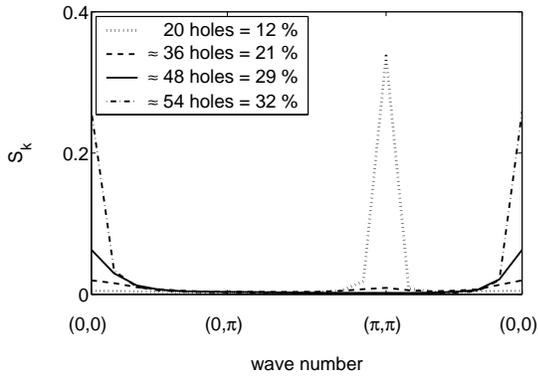}
  \caption{Spin-spin correlation in momentum space for various hole doping,
  $J'=0.02, J_\textnormal{H}=6$ and $\beta = 50$.\label{fig:ssk_MCJ_se0.02_FM.eps}}
\end{figure}

For small to medium $J' \lesssim 0.03$,
this region is ferromagnetic because of the dominant double-exchange. The corespin structure factor
$S(\vec k)$  is depicted in
fig.~\ref{fig:ssk_MCJ_se0.02_FM.eps} for $J'=0.02$ and variable filling. The
curve for $x\approx 12\%$, which is in the polaronic regime, shows a clear
AFM signal. There is a barely visible FM signal, but it is very small in
spite of the rather large doping, because
the polarons are independent and not mutually aligned. Just below $\mu^*$,
corresponding to $x\approx 21\%$, the AFM  peak
has almost vanished and there is a slightly larger FM peak at $(0,0)$. This means that the
ferromagnetism is not very strong and that even some remnant of the AFM
remains, see also the corespin correlation in real space in
Fig.~\ref{hss_ss_MC_SE0.02_pol_ps}. The lattice is however homogeneous and
the spectral density (not shown) consists
of one single tight binding band with reduced hopping strength $t\approx 0.75$ and
with broadened signals. For larger doping $x \approx 29 \%$ and $x \approx
32\%$, the spin structure factor $S(\vec k)$ shows that the ferromagnetism
grows upon increasing the hole density up towards 
half-filling of the lower Kondo band. The bandwidth then increases and the
signals in the spectral density become sharper, as fluctuations of the
corespins are suppressed by the kinetic energy.

With increasing $J'$, however, the ferromagnetism gradually becomes weaker and
for $J'=0.05$, $S(\vec k) $ does not show any signal at $(0, 0)$ for any
doping $x < 0.5$. Instead, the so called ``Flux phase'' appears around half
filling. The band structure then has a pseudogap at half-filling and differs markedly from
the tight-binding-DOS of a ferromagnetic or paramagnetic lattice. The spin structure
factor shows signals at $(0,\pi)$ and
$(\pi,0)$.\cite{Aliaga_island_2d,Agterberg_00,Yamanaka_98,DaghoferSCES}

\subsection{Phase Diagram for $\beta=50$}    \label{sec:phase_diagram}

\begin{figure}
  \centering
  \includegraphics[width=0.45\textwidth]{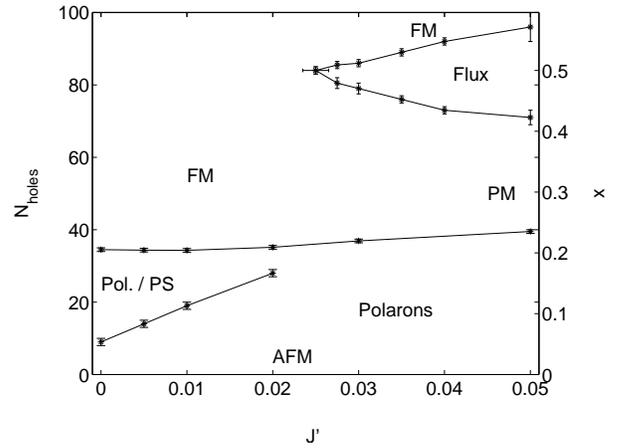}
  \caption{Phase diagram for $J_\textnormal{H} = 6$, $\beta = 50$, $0 \leq J'
  \leq 0.05$ and $0 \leq x \leq 0.6$ (i.e. filled to $40\%$ filled lower Kondo
  band) on a $14 \times 12$-lattice. ``Pol.'': polaronic regime,
  ``Pol./PS'': mixture of both polarons and larger ferromagnetic
  clusters, ``AFM'': antiferromagnet, ``FM'': ferromagnet, ``PM'': regime without magnetic
  structure, ``Flux'': Flux phase.\label{fig:phase_diagram}}
\end{figure}

In order to determine the phase diagram, shown in
Fig.~\ref{fig:phase_diagram}, criteria to distinguish the phases have
to be specified. For the crossover from the polaronic to the phase separated
regime, we choose the filling, at which the ferromagnetic regions around the
holes - detected by dressed corespin correlation \Eq{eq:hss_def} - begin to
extend, i.~e., when the nearest antiferromagnetic correlations at $r^2=5$ become
ferromagnetic. It must be emphasized that the transition is a gradual one:
larger ferromagnetic domains occasionally occur at smaller doping and above
all polarons persist to larger doping (Sec.~\ref{sec:MC_J02}). This
criterion is therefore somewhat arbitrary. 

To determine the phase boundary between the polaronic/phase separated
region and ferromagnetism on one hand, and between the polaronic regime and
paramagnetism on the other, we choose the electron density just below
the discontinuity at the critical $\mu^*$, although the
antiferromagnetic peak may still be visible at this point; the distinction is
therefore rather based on the compressibility. From there
on, a system is labeled ``ferromagnetic'' or ``paramagnetic''.

In the paramagnetic regime, we find hardly any magnetic structure for
intermediate doping, until the signals from the Flux phase begin to
appear. 
The crossover from FM to PM is continuous and depends on doping,
larger $x$ leading to more ferromagnetism. As a criterion for the transition from
FM to Flux phase, one can take the maximal change of the corespin moment with
filling $d|S_\textnormal{tot}|/dN$, because $|S_\textnormal{tot}|=0$ for the
ideal Flux phase. Obviously, this becomes less exact as the ferromagnetism
is weakened. A further indicator is the change in the spectral density, which for
the Flux phase differs very much from a tight-binding one.
The Flux phase is subject to strong finite size effects, which are
caused by the fact that only a very small part of the Brillouin zone
contributes to the few states around the pseudogap around $(\pi/2,
\pi/2)$. For this reason, lattices of different sizes, above
all square systems because of closed shell effects, may exhibit the transition
from ferro-/paramagnetism to the Flux phase at slightly different
filling. For hole dopings above the Flux phase, all systems become
ferromagnetic. 

The resulting phase diagram with the phases discussed above is shown in
Fig.~\ref{fig:phase_diagram}.

\subsection{Effects of a lower temperature}                   \label{MC:beta}

\subsubsection{$J'=0.02$: Polarons stabilized further}    \label{MC:beta:J'002}

\begin{figure}[tbp]
  \centering
  \includegraphics[width = 0.46\textwidth,trim= 0 50 0 50,clip]{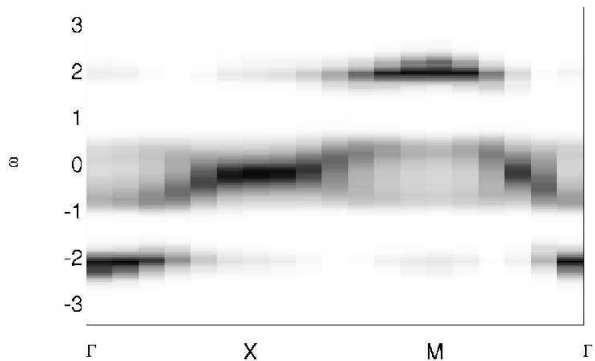}
  \caption{Spectral density (same as fig.~\ref{fig:spec_beta50_N148_02},
  but for lower temperature $\beta=80$.)  \label{fig:spec_Jse0.02_beta80_N148}}
\end{figure}

\begin{figure}
  \subfigure[]{\includegraphics[height=0.15\textwidth]
    {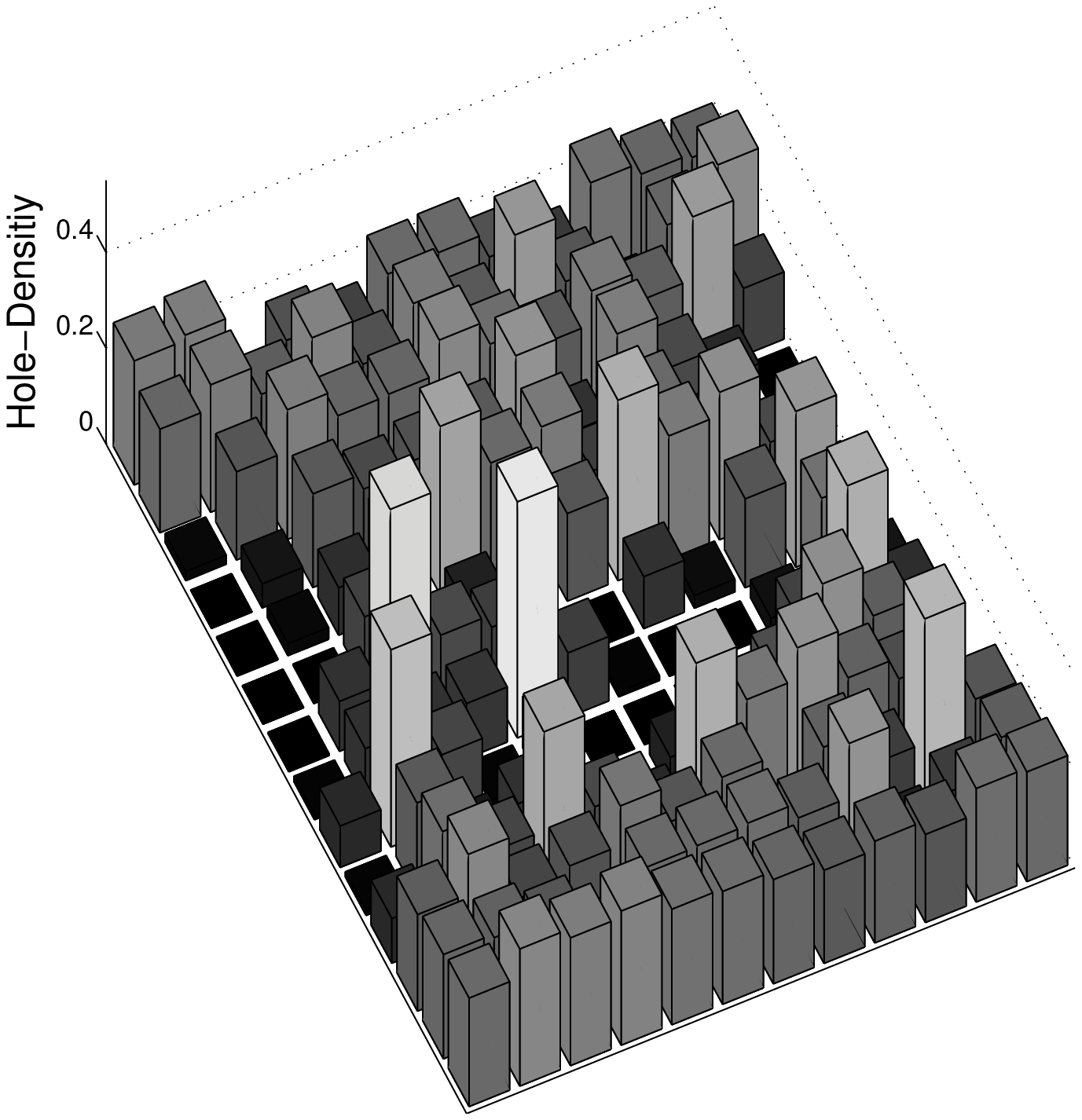}\label{MC_snp_N140_b50}}
  \subfigure[]{\includegraphics[height=0.15\textwidth]
    {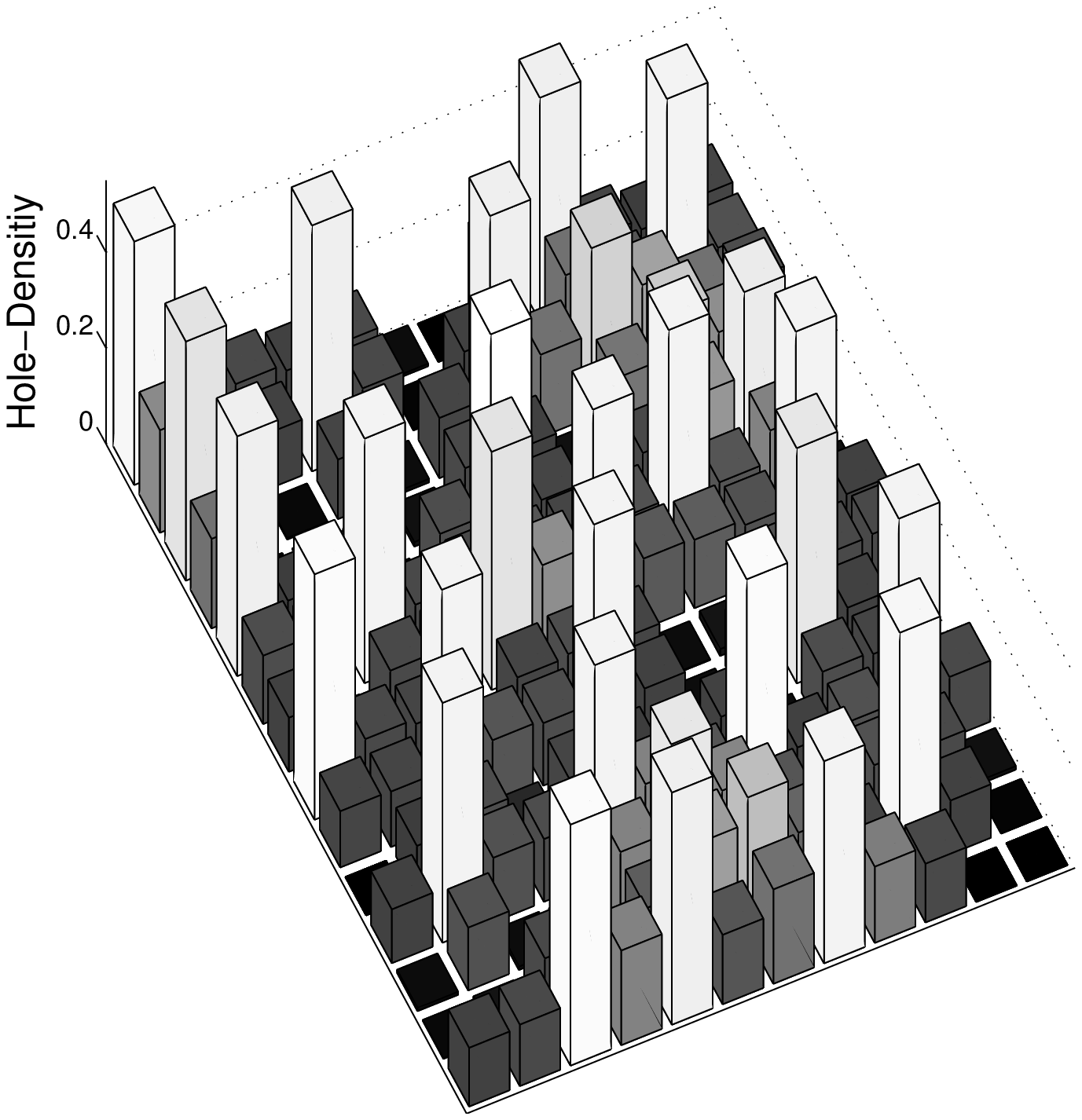}\label{MC_snp_N140_b80}}
  \subfigure[]{\includegraphics[height=0.15\textwidth]
    {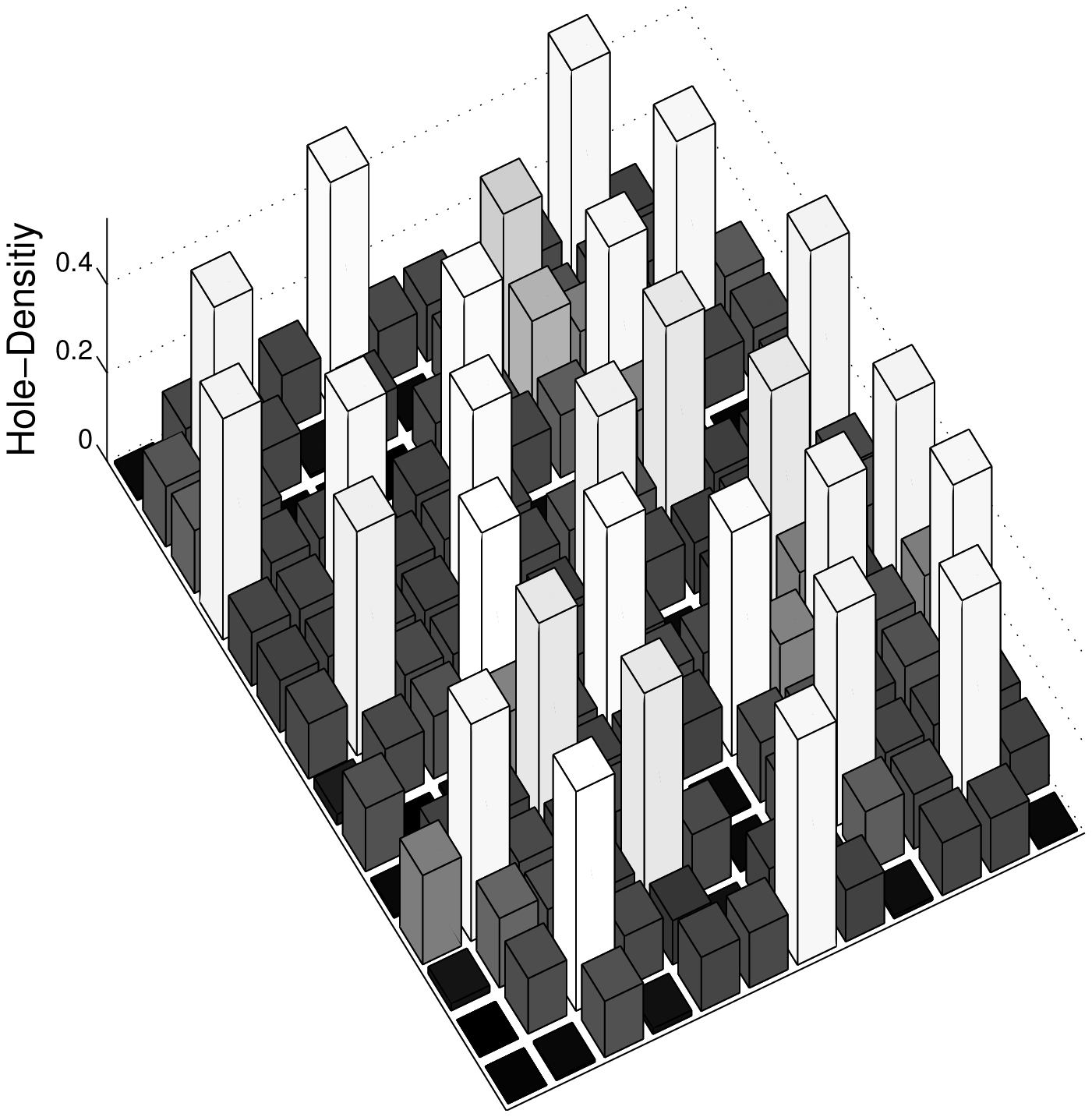}\label{MC_snp_N140_b100}}
  \caption{MC snapshot of the hole density for 28 holes ($x \approx 17\%$)
    at temperatures (a) $\beta=50$, (b) $\beta=80$, (c) and $\beta=100$.
    Other parameters as in
    Fig.~\ref{MC_snapshot_J0.02_beta80_144.eps}.\label{MC_snp_N140_b}}
\end{figure}

For $J'=0.02$, where the system can be well described by independent polarons at
$\beta=50$ and $x\lesssim 12\%$, cooling to $\beta=80$ suppresses overlapping polarons, larger ferromagnetic
regions and phase separation. 
The suppression of overlapping polarons can, e.g., be seen by comparing the spectral
density at doping $x=12\%$ (20 holes) for $J'=0.02$ and $\beta=80$
(Fig.~\ref{fig:spec_Jse0.02_beta80_N148}) to the one for $\beta=50$
(Fig.~\ref{fig:spec_beta50_N148_02}). Besides the narrowed
antiferromagnetic band, almost only one polaron signals are seen at the
lower temperature, which means that the polarons are not independent but avoid overlapping. 

The suppression of phase separation by lower temperature is clearly visible in the MC snapshots
for 28 holes ($x\approx 17\%$) depicted in Fig.~\ref{MC_snp_N140_b}. While only a few polarons exist for $\beta=50$,
they dominate at $\beta=80$, where only a few holes are delocalized, and even
more so at $\beta=100$, where two polarons hardly ever share a site.
 
As for $\beta=50$ and $J'=0.05$, the reason for the polarons' dominance over PS
is their antiferromagnetic stacking. Since each polaron takes 5 lattice
sites, they can obviously no longer be stacked in this way for dopings larger than
$20\%$. As can be seen in the phase diagram in Fig.~\ref{fig:phase_diagram_b80},
this doping is the point, where the larger clusters begin to dominate.
In this case, the polaronic phase is more ordered than a phase separated
scenario because of the stacking, and PS therefore occurs at higher temperature because of its higher entropy.

\begin{figure}
  \includegraphics[width = 0.4\textwidth]{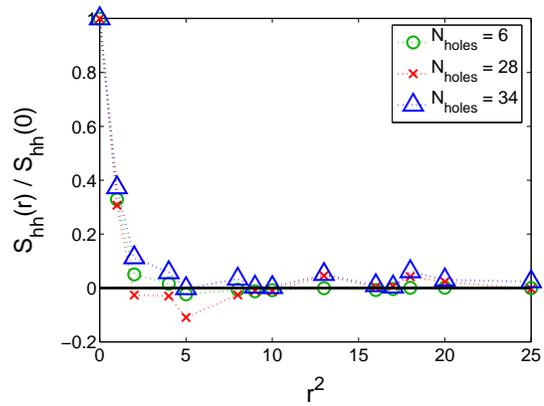}
  \caption{(Color online) Hole-Spin-Hole-Spin correlation Eq.\ref{eq:hhss_def} for
    $J'=0.02$, $\beta=80$
    and various fillings. Other
    parameters as in Fig.~\ref{fig:hss_MC00vsMC02}} 
  \label{fig:hhss_Js0.02_b80_Nvar}
\end{figure}

The stacking can be seen in the Hole-Spin-Hole-Spin correlation depicted in
Fig.~\ref{fig:hhss_Js0.02_b80_Nvar} for various doping levels. While  one
sees only short range correlations for 6 holes ($x=3.5\%$), the antiferromagnetic
stacking can be clearly observed for 28 holes ($x=17\%$). This order
disappears again for 34 holes ($x=20\%$), when the polarons begin to
merge. At this point, some tendencies to diagonal chains (see below) may
exist at very low temperatures ($\beta=100$). 

In the ferromagnetic part of the phase diagram, a spin canted phase becomes
actually more stable than ferromagnetism for some electron fillings and for
periodic boundary conditions. Its spin structure factor has a
peak for the smallest reciprocal lattice-vector of the system ($\pi/7$ for a
$14 \times 12$ lattice). Analytic energy comparison of the FM and this
canted phase shows that the range of fillings, where the canted phase has
lower energy, becomes smaller for larger lattices. Moreover, it
is not present for open boundary conditions and thus probably a finite size
effect. It shows, however, that ferromagnetism is not very dominant for doping
levels near the phase separated regime.

\subsubsection{$J'=0$: Tendency to diagonal stripes}    \label{MC:beta:J'0}

\begin{figure}
  \includegraphics[width=0.3\textwidth]{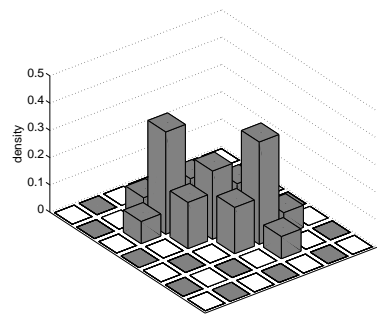}
  \caption{Corespin configuration with three flipped spins, filled with two
    holes. Details as Fig.~\ref{polaron_conf}.\label{fig:dbl_polaron_3}}
\end{figure}

\begin{figure}[htbp]
  \centering
  \subfigure[]{\includegraphics[width = 0.23\textwidth]
    {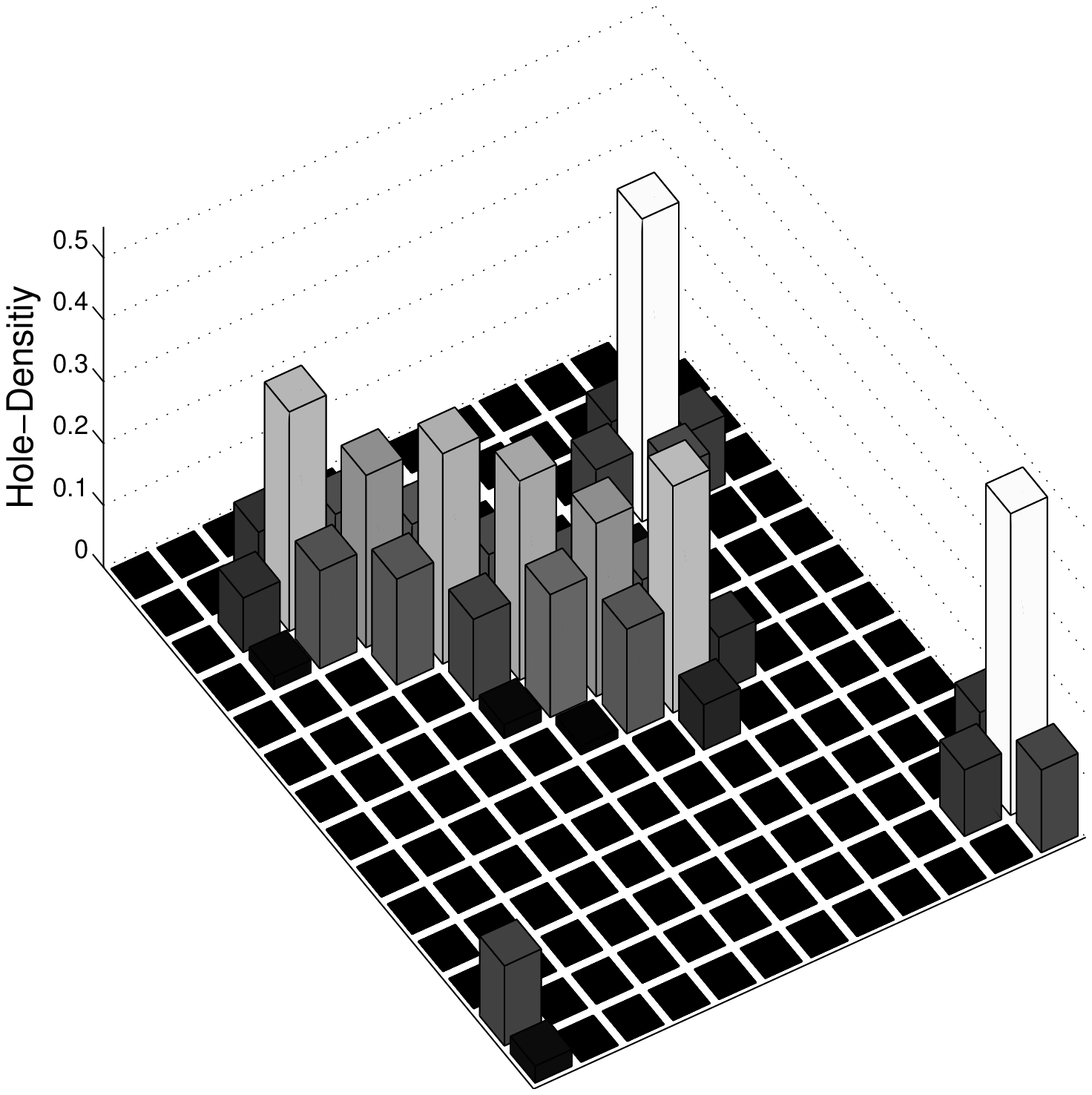}\label{fig:MC_snapshots_J0.0_beta80_162}}
  \subfigure[]{\includegraphics[width = 0.23\textwidth]
    {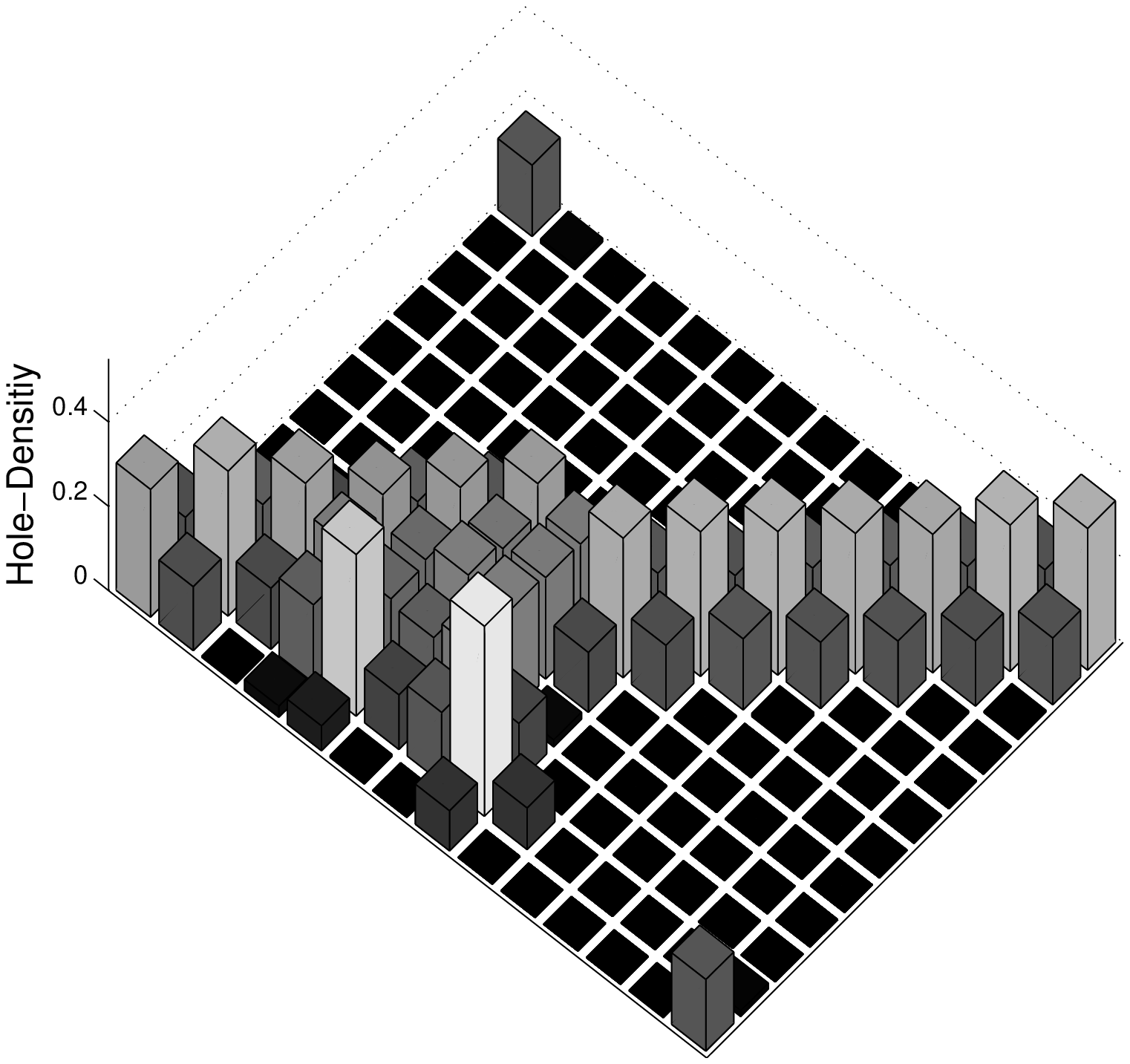}\label{fig:MC_snapshots_J0.0_beta80_156}}
  \caption{Monte Carlo snapshots for $J'=0.0$ and $\beta=80$ and (a) $6$ holes
  ($x\approx 4\%$), (b) $12$ holes ($x\approx 7\%$).\label{fig:MC_snapshots_J0.0_beta80}
  Height represents hole density, grayshades are for 
    better visibility.}
\end{figure}

\begin{figure}[htbp]
  \centering
  \subfigure{\includegraphics[width = 0.23\textwidth]{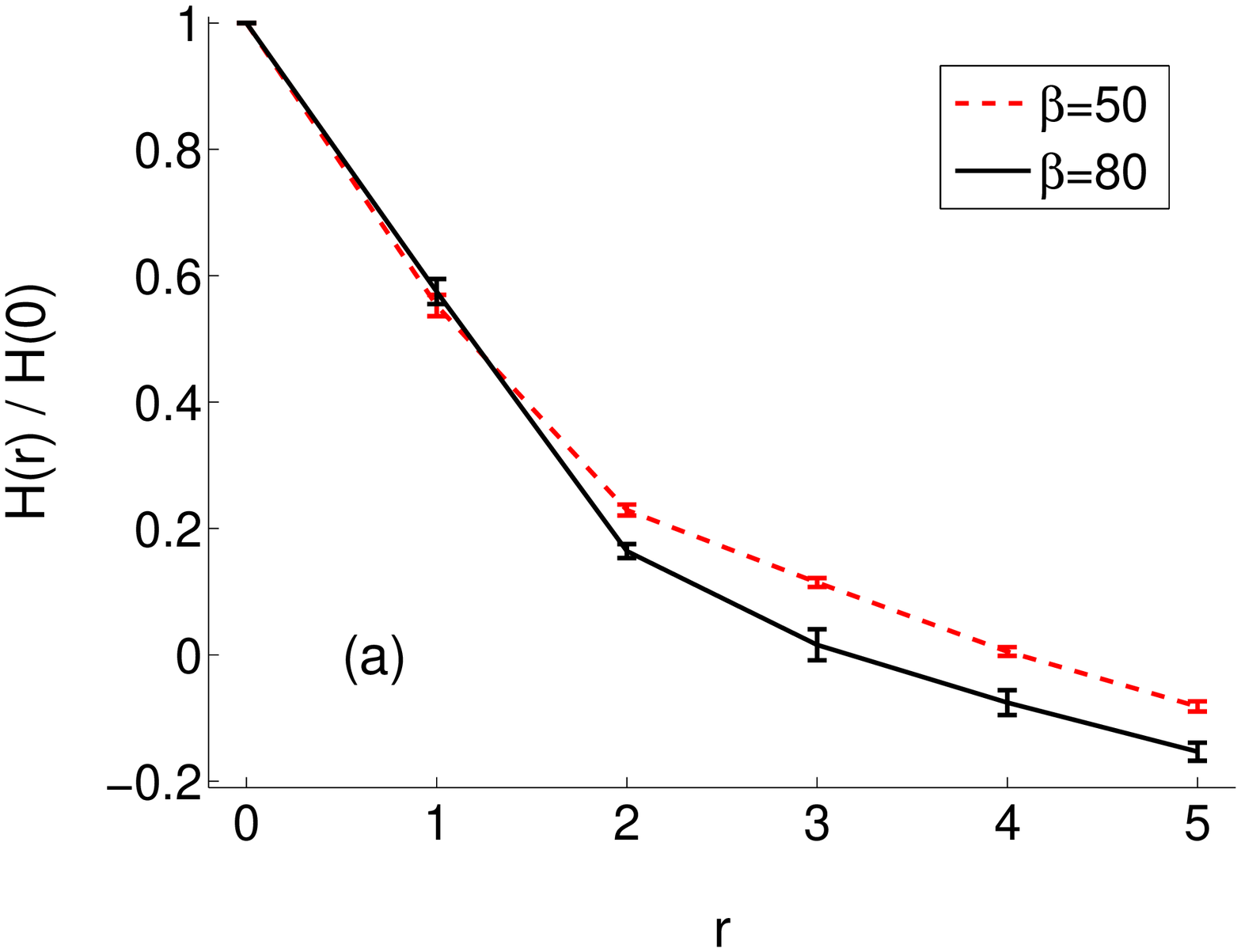}\label{fig:cc_Jse0.00_N156_str}}
  \subfigure{\includegraphics[width = 0.23\textwidth]{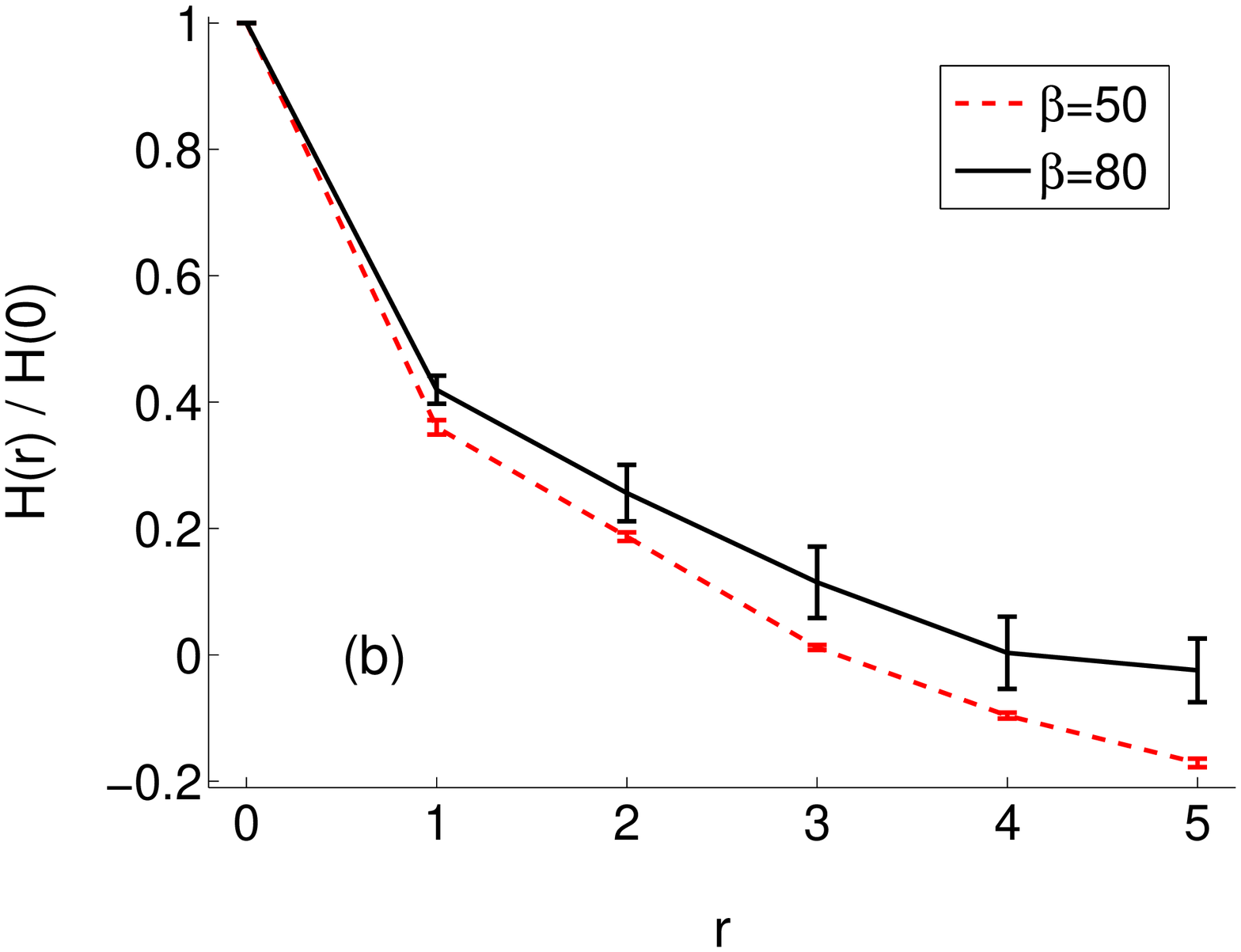}\label{fig:cc_Jse0.00_N156_dia}}
  \caption{(Color online) Hole correlation \Eq{eq:hh_def} for $J'=0.0$ and $12$ holes ($x=7.1\%$): The
  left panel (a) shows the correlations in $(1,0)$-direction; the right panel (b)
  shows the same for the diagonal $(1,1)$-direction. Dashed line: $\beta=50$,
  solid: $\beta=80$.\label{fig:cc_Jse0.00_N156}}
\end{figure}

In Ref.~\onlinecite{DaghoferKoller2003}, we showed that larger FM clusters with two
flipped spins and two holes have (slightly) higher energy than two separated
polarons. Fig.~\ref{fig:dbl_polaron_3} depicts another possibility with three
flipped spins but only two holes. The total kinetic energy of the two holes is then
$0.61\,t_0$ lower than that for two independent polarons. However, 3 spins
have to be flipped instead of two, which costs the energy $8\times
J_\textrm{eff}$. The gain and the loss balance each other for
$J_\textrm{eff}\approx 0.077$ in this simple model calculation. In the MC
simulations, we observe tendencies to such configurations for
$J'=0\;(J_\textrm{eff}\approx0.083$), which extend to 
diagonal chains at higher doping, see the snapshots in
Fig.~\ref{fig:MC_snapshots_J0.0}. For larger hole doping, isotropic clusters
coexist with these diagonal chains and finally dominate.
In order to measure these diagonal chains, we use the hole-correlation
function
\begin{equation} \label{eq:hh_def}
  H(\vec r) = \frac{1}{L} \sum_{\vec i} (n^h_{\vec i}-\bar n_h )\
  (n^h_{\vec i + \vec r}-\bar n_h)\;,
\end{equation}
where $n_{\vec i}^h$ is the expectation value of the hole density at site
$\vec i$ for a given corespin configuration and $\bar n_h$ is the overall hole
density. As explained for Eq.~\ref{eq:hhss_def},
this is not the usual density correlation function. The results for $12$
holes ($x=7.1\%$) are depicted in Fig.~\ref{fig:cc_Jse0.00_N156} and show
that diagonal correlations are slightly enhanced at $\beta=80$.

\subsubsection{$J'=0.05$: Vertical stripes}    \label{MC:beta:J'0.05}

\begin{figure}
  \subfigure[]{
    \begin{minipage}{0.15\textwidth}
      \vspace*{0.5em}
      \includegraphics[width = 0.99\textwidth]{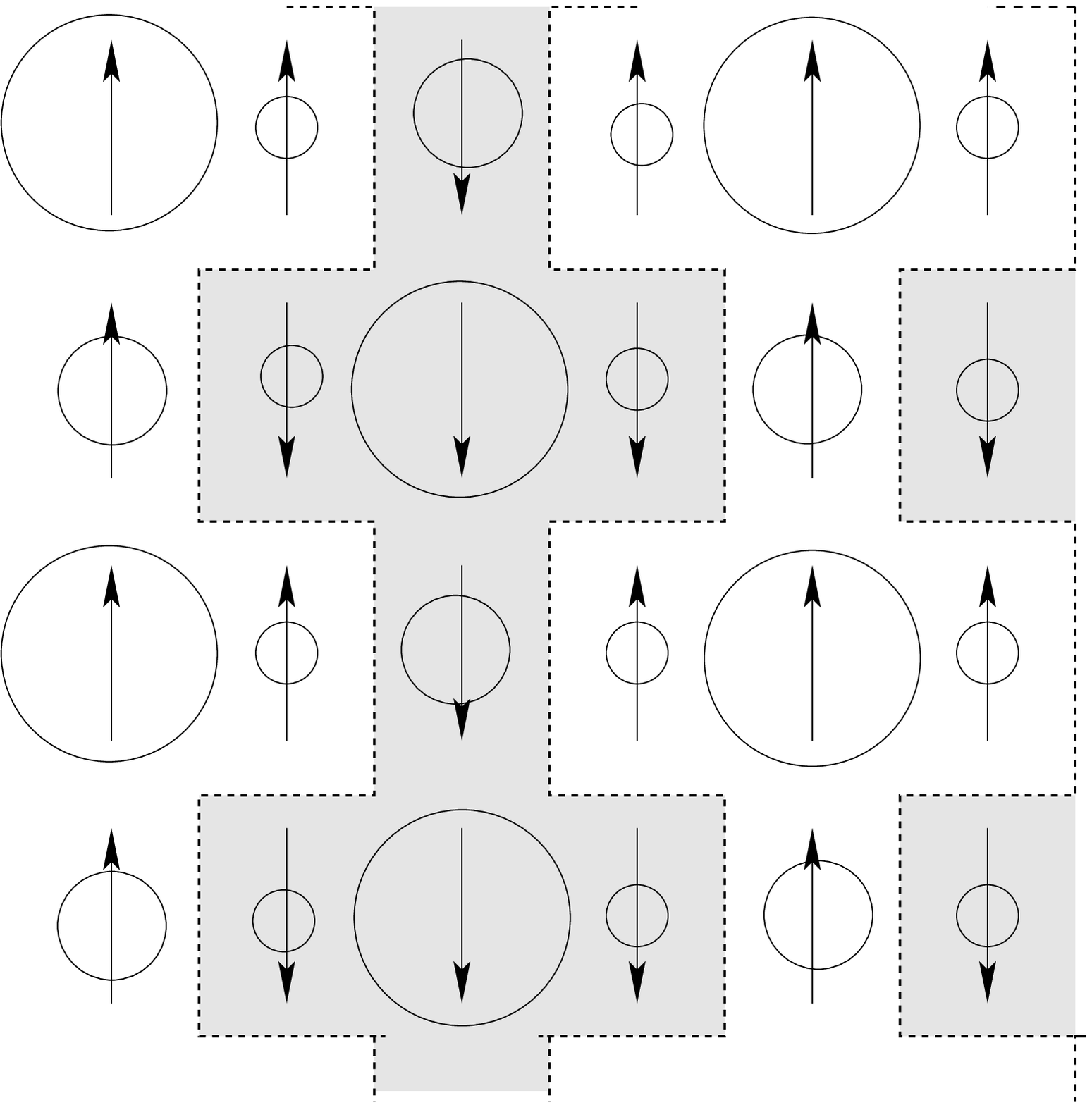}
      \vspace{1em}
    \end{minipage}\label{stripes_spins_schem}}
  \hfill
  \subfigure[]{
    \begin{minipage}{0.3\textwidth}
      \includegraphics[width = 0.99\textwidth]{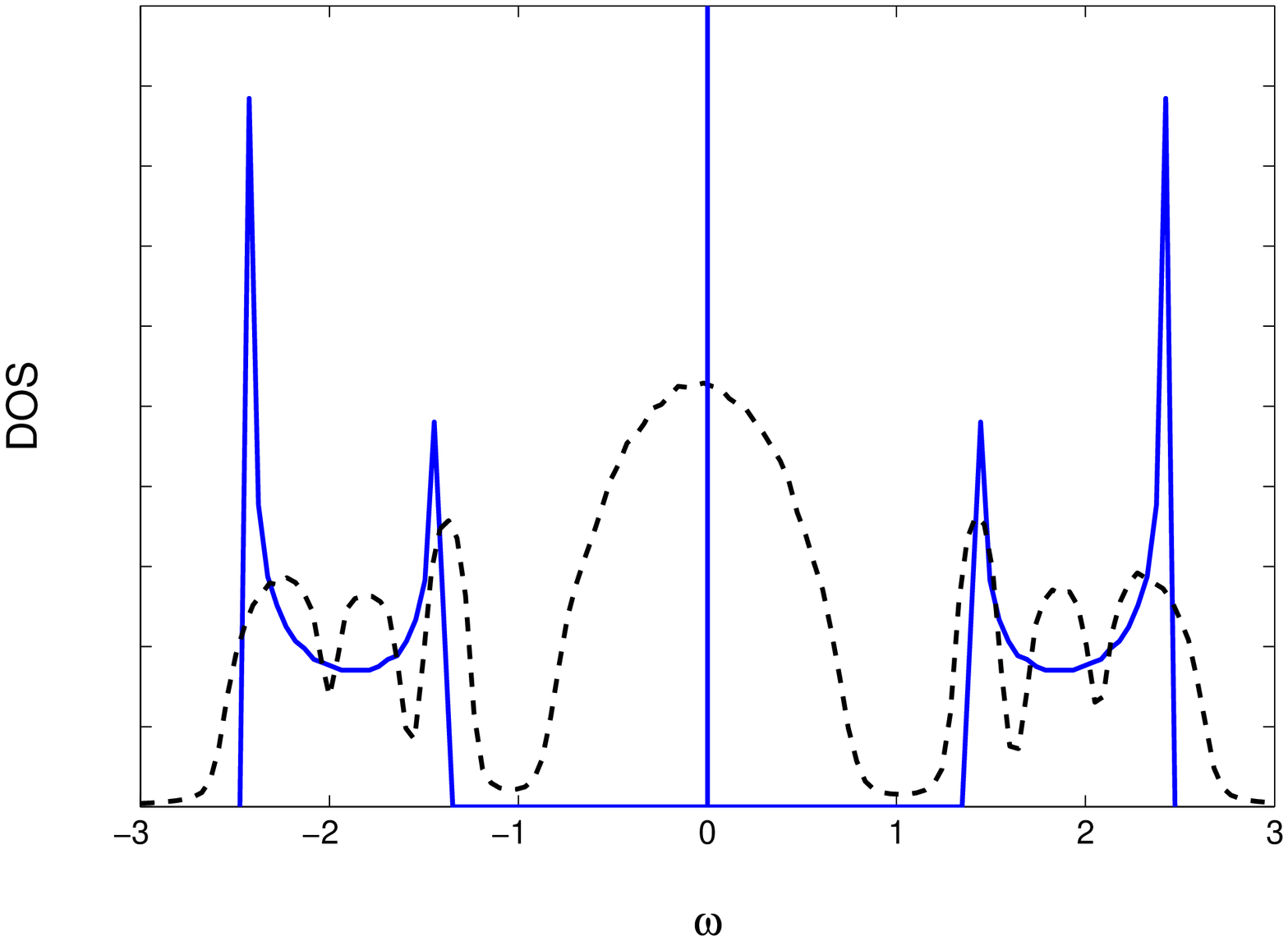}
    \end{minipage}
    \label{stripes_dos}}
    \caption{(Color online) Stripe Phase for $J'=0.05$ and $x=0.25$. Left: Schematic
    representation of the corespins. The circles represent the
    holes density, but only qualitatively. Right: one-particle density of states of
    the two dispersive bands for $J_H\to \infty$ and the thermodynamic limit
    (continuous line) and for finite $J_H$ on a $12\times 14$ lattice with random
    fluctuations added to the corespins.\label{fig:stripes_spins}}
\end{figure}

For $J'=0.05$ and doping $x=0.25$, low temperatures favor a stripe-phase with a
periodicity  of 4 sites perpendicular to the stripes. The spin
configuration is schematically depicted in
Fig.~\ref{stripes_spins_schem}. The corespins are drawn within the $xy$-plane for
visibility, but such configurations are not preferred over others with the
corespins perpendicular to the plane of the 2D-lattice, because the
Hamiltonian only depends on the relative orientation of neighboring spins,
and a global rotation therefore has no effect.

\begin{figure}
  \subfigure[]{\begin{minipage}{0.2\textwidth}
      \vspace*{-18em}
    \includegraphics[width =
    0.95\textwidth]{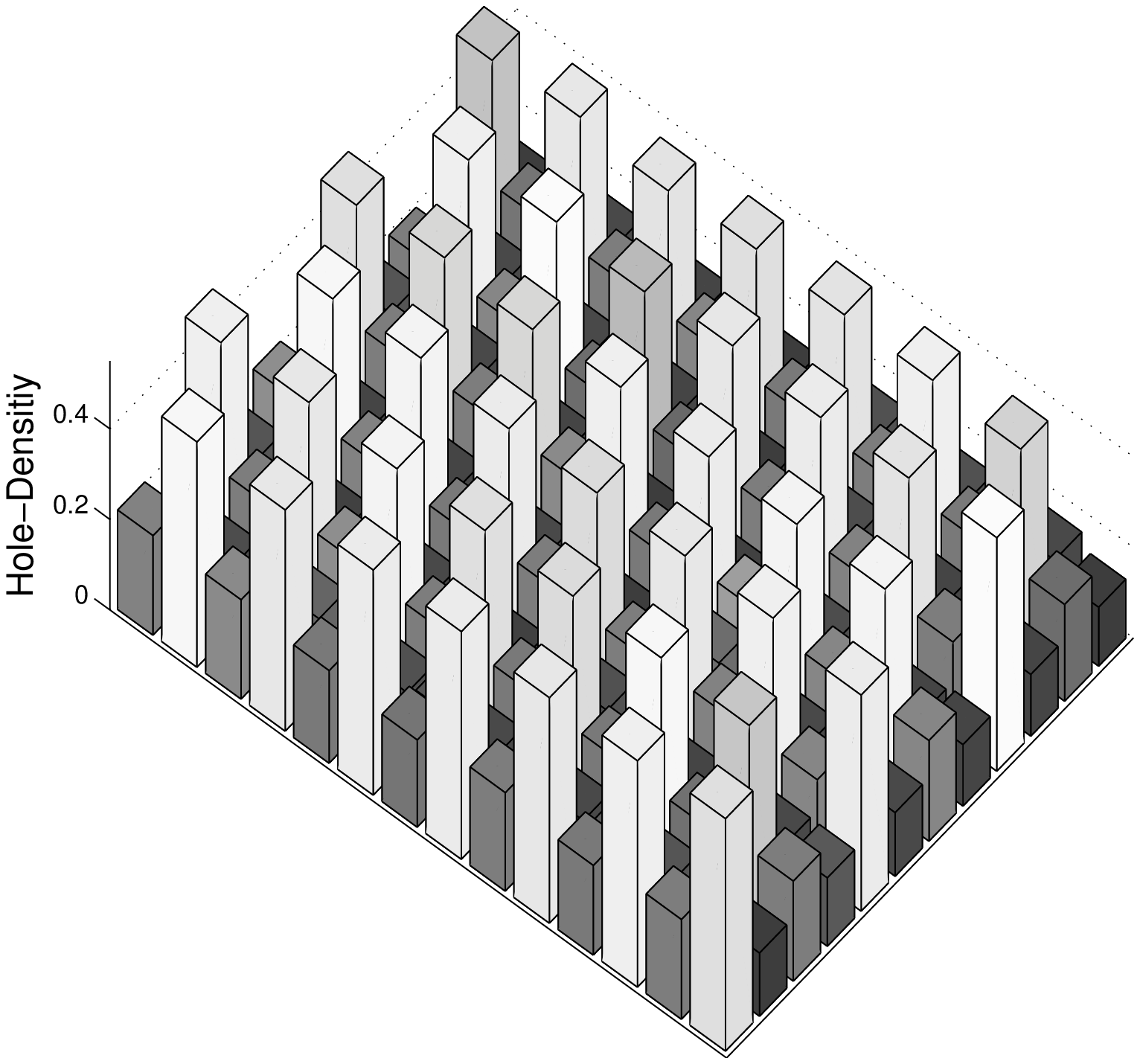}\\
    \end{minipage}
    \label{stripes_spins_snap}}
  \subfigure[]{
  \includegraphics[width = 0.25\textwidth]{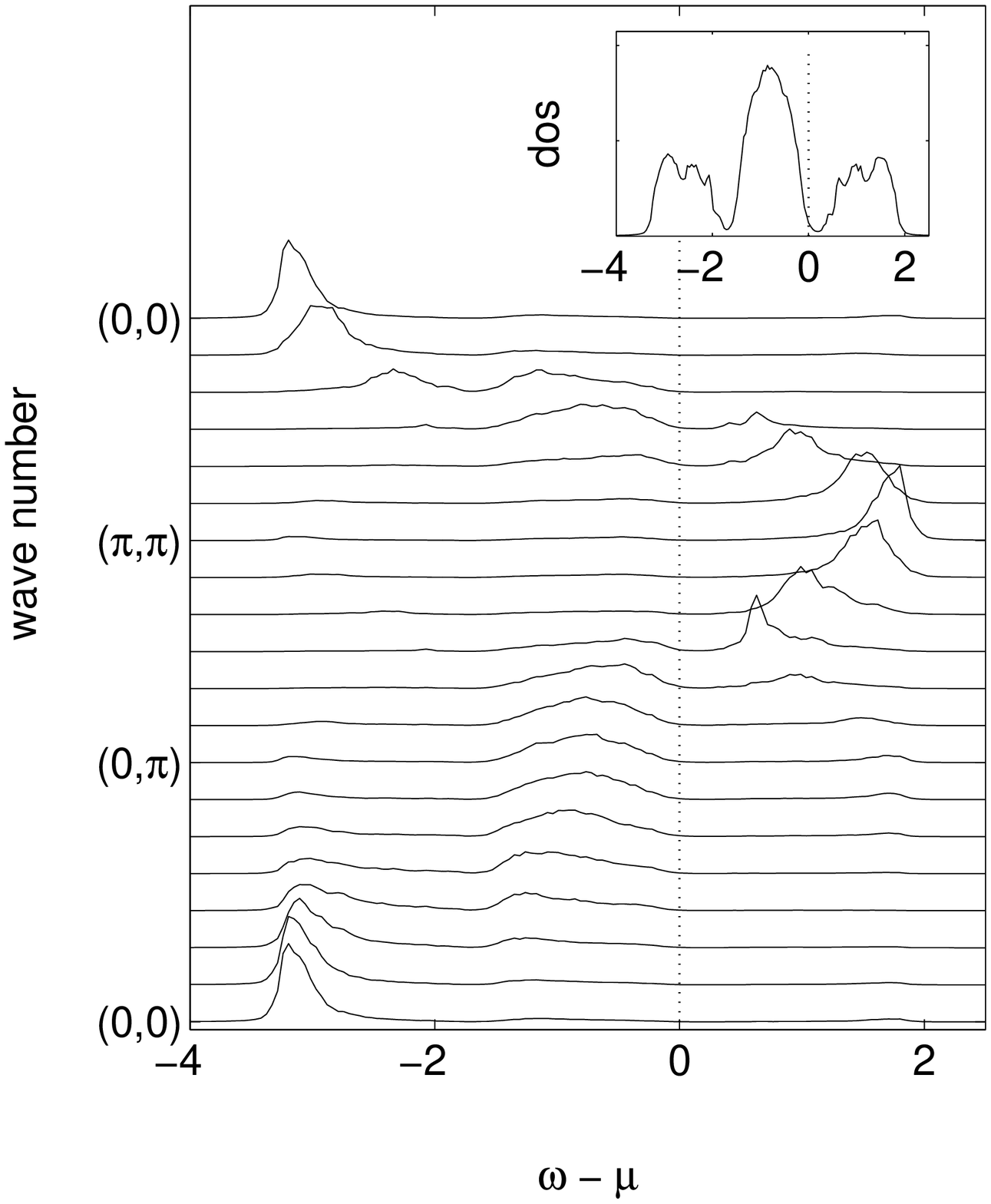} \label{fig:Ak_MC_Jse0.05_beta80_N126.eps}}
  \caption{MC snapshot and spectral density for $J'=0.05, \beta=80$, and
    $x=0.25$. Left: MC snapshot:  The heights represent hole densities,
    grayshades indicate FM (white) and AFM (black) correlations to the
    nearest neighbors. Right:  Spectral density and one-particle DOS.}
\end{figure}

For the schematic corespin configuration with perfectly aligned spins and
with the approximation $J_H\to\infty$, the band structure is gapped and consists of several
dispersionless bands at $\omega = 0$ and two bands with $\eps(k_x)= \pm
\sqrt{4\cos^2(k_x)+2}$, when the stripes run in x-direction. As no hopping is possible between the antiparallel
stripes, the band does not depend on the momentum perpendicular to the
stripes. The resulting one-particle density of states for $J_H\to\infty$ and
the thermodynamic limit can be seen in
Fig.~\ref{stripes_dos} (continuous line) and it is very similar to the DOS of one-dimensional
tight-binding bands. For comparison to the MC data (inset of
Fig.~\ref{fig:Ak_MC_Jse0.05_beta80_N126.eps}), we also calculated the
spectral density for finite $J_H$ on a $12\times14$ lattice for the model
corespin configuration with slight additional
fluctuations (dashed line in Fig.~\ref{stripes_dos}). The fluctuations were chosen so as to yield
approximately the same width for the central  band as the MC data.
On a $24\times12$ lattice, the stripes can only develop in one
direction, because 14 is not divisible by 4. One therefore can see the
dependence of these bands on only one direction of the momentum quite clearly in the
averaged spectral density depicted in
Fig.~\ref{fig:Ak_MC_Jse0.05_beta80_N126.eps}. Fig.~\ref{stripes_spins_snap}
shows an MC snapshot of this phase, where one sees stripes in the hole density.

\begin{figure}
   \includegraphics[width=0.4\textwidth]{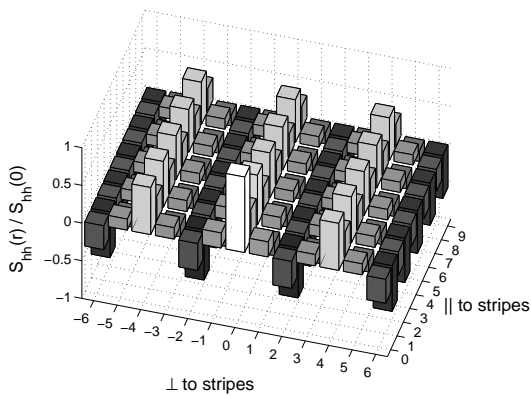}
   \caption{Hole-Spin-Hole-Spin correlation Eq.\ref{eq:hhss_def} for $J'=0.05, \beta=80$ and
     $x=0.25$ ($42$ holes). Grayshades for better visibility. \label{hh_J05_beta80_126.eps}}
\end{figure}

Figure~\ref{hh_J05_beta80_126.eps} shows the Hole-Spin-Hole-Spin correlation
function Eq.\ref{eq:hhss_def} for $x=0.25$: One clearly sees the regular
array of holes and the stripes in the corespins as
depicted in Fig.~\ref{stripes_spins_schem}.

The compressibility around this filling is very large, i.e., there is a jump
in the electron density versus the chemical potential from $x\approx
0.29$ to $x \approx 0.23$ and the filling $x=0.25$ can not be
stabilized in the grand canonical ensemble. The discontinuity is not present
for the higher temperature $\beta=50$, where there is no stripe phase. At
fillings slightly away from $x=0.25$,
stripes develop in parts of the lattice and are mixed with polarons or larger
clusters.

The occurrence of this stripe phase at $x=0.25$ and $J'=0.05$ is in contrast to
the results reported by Aliaga {\it et al.}~\cite{Aliaga_island_2d}, who found
an island phase of antiferromagnetically stacked FM clusters of size 2 by
2 for this filling. The electron density is homogeneous in this phase, and
every spin has two FM and two AFM neighbors. Their calculations were done for the $J_H\to \infty$
model, but with a larger $J'$, so as to give approximately the same
effective parameter $J_\textrm{eff}$. This island phase can however not be
stable on a $12\times14$ lattice. In order to check the validity of our
results, we therefore repeated the simulations on a $12\times12$ system and still
found stripes.
Numeric comparison of the ground-state energies for perfect stripe and
perfect island phase on a $12\times 12$ lattice gave a lower energy
for stripes for $J_H=6$ and $J'=0.05$ (corresponding to
$J_\textrm{eff}\approx 0.133$ for $x=0$ and to $J_\textrm{eff}\approx
0.1125$ for $x=0.25$), but favored the island phase for $J_H=\infty$
and  $J'=0.133$ or $J'=0.1125$. The reason lies in the density
dependence of the AFM coupling coming from the virtual excitations
(second term in Eq.~\ref{eq:H}), because the sites with 3 AFM and only
one FM neighbor have high electron density and therefore maximize the
gain from  this interaction, whereas the holes weaken the interaction
on the sites with predominately FM neighbors. This therefore presents
a feature which is missing from the first order $J_H\to \infty$
approximation but is included in the second order approach to the FM
Kondo model. 

\subsubsection{Phase diagram at $\beta=80$}    \label{MC:beta:phase_diagram}

\begin{figure}
  \centering
  \includegraphics[width=0.45\textwidth]{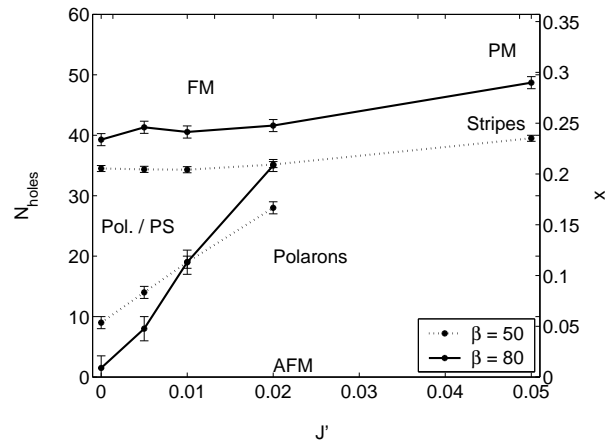}
  \caption{Phase diagram for $\beta=80$ and $\beta=50$. Similar as
  fig.~\ref{fig:phase_diagram}, but for a smaller range of doping. Solid 
  lines: results for $\beta=80$, dashed lines: $\beta=50$.\label{fig:phase_diagram_b80}}
\end{figure}

Our findings are summarized in the phase diagram for $\beta = 80$,
$0<x\leq0.36$ and $ 0 \leq J' \leq 0.05$ depicted in
Fig.~\ref{fig:phase_diagram_b80}. It shows that lower temperatures lead to
larger clusters and PS for small $J'<0.01$, while they favor individual polarons
for more realistic values $J'>0.01$. For the lower temperature, the homogeneous ferro-/para-magnetic phase
only sets in at higher doping.

\section{MC Results for 3D}                 \label{sec:Numerical_Results_3d}
\begin{figure}[ht]
  \includegraphics[width=0.47\textwidth]{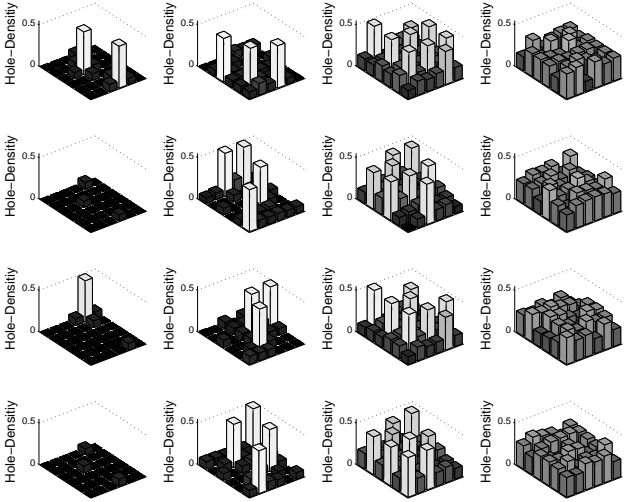}\\
  \caption{MC snapshots for (a) 3
    holes ($x\approx 2\%$), (b) 14 holes ($x\approx 10\%$), (c) 32 holes ($x\approx 22\%$)
    and (d) $\approx 38$ holes ($x\approx 26\%$). The four plots stacked one
    above the other represent the 4 layers of the $6 \times 6 \times 4$
    lattice, height represents hole density, grayshades are for
    better visibility. $J_H=6, \beta=50, J'=0.02$.}
  \label{MC_snap_3d}
\end{figure}

In three dimensions, we use a lattice of $6 \times 6 \times 4$ sites with
periodic boundary conditions. For selected doping levels, we have
repeated the simulations on lattices of different size in order to
control finite size effects.
As in one and two dimensions, the corespins are antiferromagnetically
aligned for the filled lower Kondo band and the system exhibits a
transition to ferromagnetism upon doping.
Figure~\ref{MC_snap_3d} shows MC snapshots for $J'=0.02$ and
$\beta=50$ at various doping levels. For three holes ($x\approx 2\%$), one 
clearly sees three polarons of 7 sites each. The polarons also remain well
separated for 14 holes ($x\approx 10\%$), although they take up $\approx 68\%$ of the lattice
at this doping level. As every polaron contains 7 sites, no more than
$x\approx 14\%$ holes can be doped into the lattice in this way. For the larger
doping level $x\approx 22\%$ (32 holes), one sees tendencies toward  a
superstructure of regularly arrayed holes, see Fig.~\ref{MC_snap_3d}(c), which we termed `polaron lattice'.
\begin{figure}[ht]
   \includegraphics[width=0.2\textwidth]{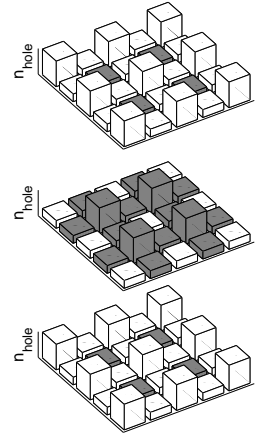}
   \caption{Schematic representation of the corespins
   and density for the `polaron lattice'. Empty(filled) spins denote spin
     up(down), height represents hole density.
     \label{pol_latt_3d}}
 \end{figure}
An idealized representation of the polaron lattice is depicted in
figure \ref{pol_latt_3d}, which is for a doping concentration of
$x=0.25$. The idealized version consists of two sublattices with
opposing corespin and hole density is high on sites with four FM
neighbors.  
Interestingly, at this 'optimal' doping concentration for the polaron
lattice, the actual MC simulations favor a more homogeneous and
disordered structure, like the one depicted in Fig.~\ref{MC_snap_3d}(d), which
has been obtained for $x\approx 27\%$ ($\approx 39$ holes). 

On very small systems ($4\times4\times4$ and $4\times4\times6$), the
polaron lattice competes with a three-dimensional form of the vertical
stripes found in 2D for $J'=0.05$ and $\beta=80$
(Sec.~\ref{MC:beta:J'0.05}). Due to their periodicity, the stripes can
only develop on lattices, where at least two of the three dimensions
are divisible by four, and they do therefore not occur on a
$6\times6\times4$-system. Numeric comparison of the energies of the
polaron lattice and the stripe-structure for $x=0.25$
on systems of different size revealed that the polaron lattice has lower energy
for systems which are larger than 4 in at least two directions, i.e., for
lattices which are not of the shape $4\times 4\times L_z$. In order to verify
this result, we performed Monte Carlo simulations on an $8\times6\times4$-system
and indeed found a polaron lattice.

\begin{figure}[ht]
\includegraphics[width=0.35\textwidth]{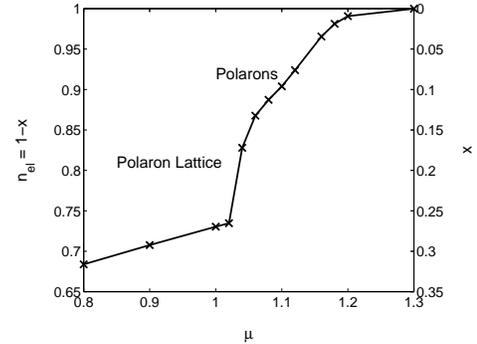}
\caption{Electron density $n_\textrm{el} = 1-x$ vs. chemical potential $\mu$
  in 3D. Parameters as Fig.~\ref{MC_snap_3d}.\label{mu_N_3d_02}} 
\end{figure}

The dependence of the electron density on the chemical potential is given in
Fig.~\ref{mu_N_3d_02}. The slope of the curve $n_\textrm{el}$ vs. $\mu$
(i.\,e. the compressibility) is finite for the polaronic phase near
$n_\textrm{el}=1$ in contrast to the results for independent
polarons,\cite{DaghoferKoller2003,KollerPruell2002c} because the polarons avoid overlapping and
it therefore becomes harder to fit more holes into the lattice. At
$x\approx1/7$ ($n_\textrm{el}\approx0.86$), when there is no more room for
polarons, the situation changes. The slope of $n_\textrm{el}$ vs. $\mu$
becomes steeper, and the polaron lattice  partially develops until
$x\lesssim 23\%$, where the homogeneous disordered phase sets
in. At $x \approx 27\%$, the compressibility is suddenly much reduced. It is
not clear, why the homogeneous phase has such a high compressibility for
$0.23<x<0.27$, and this may be a finite size effect. The lattice is not phase
separated in this regime, snapshots are similar to the one depicted in
Fig.~\ref{MC_snap_3d}(d), with some sites (scattered throughout the lattice)
having much higher or lower density than others. After ferromagnetism has
become established at still higher doping values ($x \gtrsim 30\%$), the density
becomes more evenly distributed (not shown). 

\begin{figure}[ht]
  \includegraphics[width=0.4\textwidth]{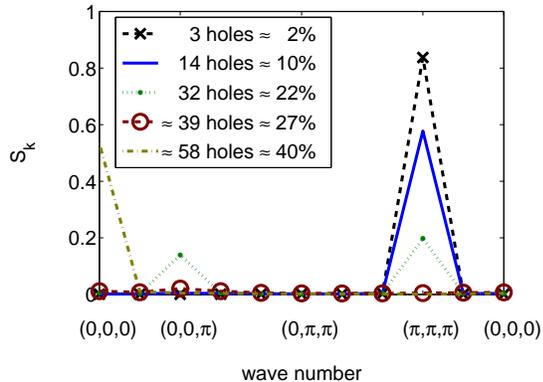}
  \caption{(Color online) 3d Corespin Structure Factor for various doping levels. Parameters
    as Fig.~\ref{MC_snap_3d}\label{ssk_3D_J_se0.02}}
\end{figure}

Figure \ref{ssk_3D_J_se0.02} shows the momentum
dependent corespin correlations for approximately the same dopings as the
snapshots. For the almost filled band (3 holes, $x\approx 2\%$), it shows antiferromagnetism which
decreases slightly for 14 holes ($x\approx 10\%$). For 32 holes ($x\approx 22\%$), the partial polaron lattice, a
second peak appears for $k=(0, 0, \pi)$. 
For the polaron lattice, according to figure \ref{pol_latt_3d}, the majority
spin for consecutive planes alternates, which explains the 
signal at $(0,0,\pi)$ in $S(\vec k)$. The corespin structure is isotropic, the same
alternating structure occurs in $x-$ and $y-$ direction and analogous signals
appear at $(0,\pi,0)$ and $(\pi,0,0)$.
For the more homogeneous phase at $\approx 39$ holes there is no magnetic
structure, and ferromagnetism only begins to set in for $\approx 45$ holes
($x \approx 30\%$, not shown) and it is established for $\approx 58$
holes ($x \approx 40\%$).

\begin{figure}
\includegraphics[width=0.4\textwidth]{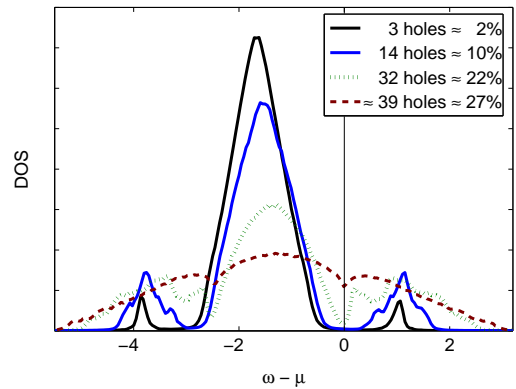}
\caption{(Color online) One-particle DOS for various doping levels. Parameters
    as in Fig.~\ref{MC_snap_3d}\label{dos_3D_J_se0.02}}
\end{figure}

Figure \ref{dos_3D_J_se0.02} shows the one-particle density of states for
various doping levels. For 3 holes ($x\approx 2\%$), one sees the broadened central band from
the AFM background and the signals from the polaronic states at $\omega\pm
\sqrt{6}$. For 14 holes ($x\approx 10\%$), these signals are stronger and somewhat broader,
because the polarons now cover a large part of the lattice and occasionally
connect. For 32 holes ($x\approx 22\%$), the
polaronic states have a considerable width, but the pseudogap is well
preserved. In will be shown later, that these bands are the dispersive bands
of the `polaron lattice'. The pseudogap has closed after the onset of the homogeneous phase at
$\approx 39$ holes ($x\approx 27\%$), but some remnant of it is still visible. For still
higher doping of $\approx 58$ holes ($x \approx 40\%$), where the lattice is
ferromagnetic, the band (not shown) is a tight-binding band with a slightly reduced
hopping amplitude.

In closing, let us discuss band structure and DOS for the ideal polaron
lattice, depicted in figure \ref{pol_latt_3d}. 
The unit cell in this case contains eight sites, and for perfectly FM/AFM corespins, the
one-particle DOS has a degenerate non-dispersive band at $\omega=-2/J_H$ and two
momentum dependent bands at energies
\begin{multline}\label{pol_latt_band}
  \eps(k_x,k_y,k_z) = - \frac{1}{J_H}\\
  \pm \sqrt{\frac{1}{J_H^2} +
  4(\cos^2(k_x) + \cos^2(k_y) + \cos^2(k_z))}\,.
\end{multline}

\begin{figure}
   \includegraphics[width=0.4\textwidth]{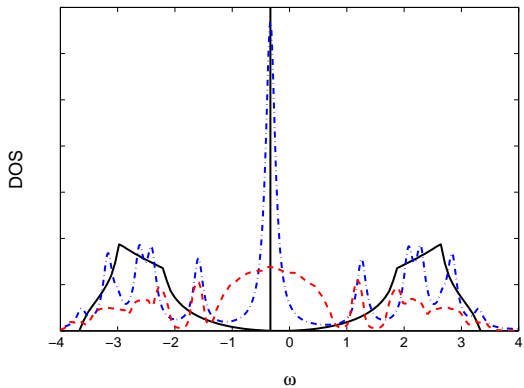}
   \caption{(Color online) One particle DOS for the ideal polaron lattice in
     the thermodynamic limit (solid), for the ideal case on a $6\times6\times4$ lattice
     (dash-dotted line), and for $6\times6\times4$ with random fluctuations around the perfect
     spin arrangement (dashed line).
     \label{dos_pol_latt_3d}}
\end{figure}

The one-particle density of states resulting from this dispersion is depicted
in Fig.~\ref{dos_pol_latt_3d} for a $6 \times 6\times 4$ lattice (dash-dotted
line). The two
dispersive bands consist of a number of peaks because of finite size
effects. For an infinite lattice (solid line), the DOS has a pseudogap
between the dispersive bands and the central $\delta$-peak.
In addition, we have performed a model calculation, in which the corespins of
the ideal polaron lattice are randomly disturbed. 
The DOS of the resulting tight-binding model with transfer integrals given by Eq.~\ref{eq:modihop} is
compared to the result of the ideal polaron lattice in Fig.~\ref{dos_pol_latt_3d}.
The amount of fluctuations has been chosen so as to give approximately the same width
for the central band as the MC data (similar to Fig.~\ref{stripes_dos}).

While the model calculation (for $x=0.25$) differs slightly
from the MC data for $32$ holes ($x\approx 22\%$) depicted in
Fig.~\ref{dos_3D_J_se0.02}, the width of the bands and their apparent
double-peak structure is reproduced. The remaining differences are probably
due to the fact that the MC data are for $x=0.22$ and the polaron
lattice is only partially developed, see the MC snapshot Fig.~\ref{MC_snap_3d}(c).

For $J'=0$, diagonally arranged polarons similar to those observed in 2D
(Sec.~\ref{MC:beta:J'0}) were found for $x\lesssim0.25$ on a $6 \times 6
\times 
4$ lattice. These stripes occurred on every second plane along the
$z-$direction (with $L_z=4$). However, this phase was \emph{not} observed on
a $6 \times 6 \times 6$ lattice and we therefore consider it a finite size
effect. On this larger lattice, we observed the `polaron lattice' instead and
the results were generally very similar to those for $J'=0.02$.

\section{Conclusions}                           \label{sec:conclusion}

We have used Monte Carlo simulations to examine the two- and
three-dimensional ferromagnetic Kondo lattice model at small to medium
hole doping and for parameters relevant to manganites.
In three dimensions, we find that a small additional antiferromagnetic
exchange $J'=0.02$ yields very similar results as $J'=0$.
No phase separation is observed at any doping level. Instead, there
are again polarons consisting of a single flipped spin in an AFM
background and containing a single hole, which form a regular `polaron
lattice' when they become very dense.

In two dimensions, we have focused on the polaronic phase and its
boundaries toward phase separation. At small hole doping, we find
polarons for all values of the superexchange $J'$.
For larger doping and small $J'$, e.g., $J'=0$ instead of $J'=0.02$,
the polarons attract each other and tend to form larger clusters.
Eventually, upon higher hole doping, phase separation into large
ferromagnetic and antiferromagnetic regions sets in.
A large antiferromagnetic superexchange ($J' = 0.05$), on the other
hand, suppresses overlapping polarons and larger ferromagnetic
regions, because it stabilizes the AFM background and favors
antiferromagnetic stacking of the individual polarons.
In the doping range $0.2 \lesssim x < 0.5$, $J'$ suppresses
the FM long-range order, so that the system is in reality rather
paramagnetic than ferromagnetic for $J'\gtrsim0.03$.

We find that lowering the temperature generally moves the homogeneous
ferro-/para-magnetic phase to higher doping.
Phase diagrams for $\beta=50$ and $\beta=80$ are given in
Fig.~\ref{fig:phase_diagram} and Fig.~\ref{fig:phase_diagram_b80}. 
A lower temperature stabilizes the polarons and disfavors larger
clusters for intermediate and large $J'\gtrsim 0.01$; for $J'=0.02$,
phase separation only occurs, when no more polarons can be fit into
the lattice.
PS is favored by lower temperatures only for unphysically small $J'
\lesssim 0.01$.
This is corroborated by a comparison of ground-state energies for idealized
phase separation and polaron scenarios, which gives phase separation for
small $J_\textrm{eff}\lesssim 0.08$ (corresponding to $J' \approx 0$ for
$J_H=6$), and polarons for larger $J_\textrm{eff}\gtrsim 0.08$.
For $J'\approx 0$, lower temperatures enhance ferromagnetic diagonal
chains with delocalized holes, while they lead to vertical stripes for
large $J'=0.05$ at doping $x=0.25$. 

\section{Acknowledgment}

This work has been supported by the Austrian Science Fund (FWF), project
no.\ P15834-PHY. We wish to thank the EPSRC (Grant GR/S18571/01) for financial
support.


\begin{thebibliography}{40}
\expandafter\ifx\csname natexlab\endcsname\relax\def\natexlab#1{#1}\fi
\expandafter\ifx\csname bibnamefont\endcsname\relax
  \def\bibnamefont#1{#1}\fi
\expandafter\ifx\csname bibfnamefont\endcsname\relax
  \def\bibfnamefont#1{#1}\fi
\expandafter\ifx\csname citenamefont\endcsname\relax
  \def\citenamefont#1{#1}\fi
\expandafter\ifx\csname url\endcsname\relax
  \def\url#1{\texttt{#1}}\fi
\expandafter\ifx\csname urlprefix\endcsname\relax\def\urlprefix{URL }\fi
\providecommand{\bibinfo}[2]{#2}
\providecommand{\eprint}[2][]{\url{#2}}

\bibitem[{\citenamefont{Kaplan and Mahanti}(1998)}]{proceedings98}
\bibinfo{author}{\bibfnamefont{T.}~\bibnamefont{Kaplan}} \bibnamefont{and}
  \bibinfo{author}{\bibfnamefont{S.}~\bibnamefont{Mahanti}},
  \emph{\bibinfo{title}{Physics of Manganites}} (\bibinfo{publisher}{Kluwer
  Academic/ Plenum Publishers}, \bibinfo{address}{New York, Boston, Dordrecht,
  London, Moscow}, \bibinfo{year}{1998}), \bibinfo{edition}{1st} ed.

\bibitem[{\citenamefont{Nagaev}(2002)}]{Nagaev:book}
\bibinfo{author}{\bibfnamefont{E.~L.} \bibnamefont{Nagaev}},
  \emph{\bibinfo{title}{Colossal Magnetoresistance and Phase Separation in
  Magnetic Semiconductors}} (\bibinfo{publisher}{Imperial College Press},
  \bibinfo{address}{London}, \bibinfo{year}{2002}), \bibinfo{edition}{1st} ed.

\bibitem[{\citenamefont{Zener}(1951)}]{zener51}
\bibinfo{author}{\bibfnamefont{C.}~\bibnamefont{Zener}},
  \bibinfo{journal}{Phys. Rev.} \textbf{\bibinfo{volume}{82}},
  \bibinfo{pages}{403} (\bibinfo{year}{1951}).

\bibitem[{\citenamefont{de~Gennes}(1960)}]{gennes60}
\bibinfo{author}{\bibfnamefont{P.-G.} \bibnamefont{de~Gennes}},
  \bibinfo{journal}{Phys. Rev.} \textbf{\bibinfo{volume}{118}},
  \bibinfo{pages}{141} (\bibinfo{year}{1960}).

\bibitem[{\citenamefont{Dagotto et~al.}(1998)\citenamefont{Dagotto, Yunoki,
  Malvezzi, Moreo, Hu, Capponi, Poilblanc, and
  Furukawa}}]{dagotto98:_ferrom_kondo_model_mangan}
\bibinfo{author}{\bibfnamefont{E.}~\bibnamefont{Dagotto}},
  \bibinfo{author}{\bibfnamefont{S.}~\bibnamefont{Yunoki}},
  \bibinfo{author}{\bibfnamefont{A.~L.} \bibnamefont{Malvezzi}},
  \bibinfo{author}{\bibfnamefont{A.}~\bibnamefont{Moreo}},
  \bibinfo{author}{\bibfnamefont{J.}~\bibnamefont{Hu}},
  \bibinfo{author}{\bibfnamefont{S.}~\bibnamefont{Capponi}},
  \bibinfo{author}{\bibfnamefont{D.}~\bibnamefont{Poilblanc}},
  \bibnamefont{and} \bibinfo{author}{\bibfnamefont{N.}~\bibnamefont{Furukawa}},
  \bibinfo{journal}{Phys. Rev. B} \textbf{\bibinfo{volume}{58}},
  \bibinfo{pages}{6414} (\bibinfo{year}{1998}).

\bibitem[{\citenamefont{Furukawa}(1998)}]{furukawa98}
\bibinfo{author}{\bibfnamefont{N.}~\bibnamefont{Furukawa}},
  \emph{\bibinfo{title}{in: Physics of manganites}} (\bibinfo{publisher}{Kluwer
  Academic Publisher}, \bibinfo{address}{New York}, \bibinfo{year}{1998}),
  \bibinfo{edition}{1st} ed.

\bibitem[{\citenamefont{Edwards}(2002)}]{EdwardsI}
\bibinfo{author}{\bibfnamefont{D.~M.} \bibnamefont{Edwards}},
  \bibinfo{journal}{Adv. Phys.} \textbf{\bibinfo{volume}{51}},
  \bibinfo{pages}{1259} (\bibinfo{year}{2002}).

\bibitem[{\citenamefont{Meyer et~al.}(2001)\citenamefont{Meyer, Santos, and
  Nolting}}]{Nolting01}
\bibinfo{author}{\bibfnamefont{D.}~\bibnamefont{Meyer}},
  \bibinfo{author}{\bibfnamefont{C.}~\bibnamefont{Santos}}, \bibnamefont{and}
  \bibinfo{author}{\bibfnamefont{W.}~\bibnamefont{Nolting}},
  \bibinfo{journal}{J. Phys. Condens. Matter} \textbf{\bibinfo{volume}{13}},
  \bibinfo{pages}{2531} (\bibinfo{year}{2001}).

\bibitem[{\citenamefont{M\"uller and Nolting}(2002)}]{Nolting03}
\bibinfo{author}{\bibfnamefont{W.}~\bibnamefont{M\"uller}} \bibnamefont{and}
  \bibinfo{author}{\bibfnamefont{W.}~\bibnamefont{Nolting}},
  \bibinfo{journal}{Phys. Rev. B} \textbf{\bibinfo{volume}{66}},
  \bibinfo{pages}{085205} (\bibinfo{year}{2002}).

\bibitem[{\citenamefont{Koller et~al.}(2002)\citenamefont{Koller, Pr\"ull,
  Evertz, and von~der Linden}}]{KollerPruell2002a}
\bibinfo{author}{\bibfnamefont{W.}~\bibnamefont{Koller}},
  \bibinfo{author}{\bibfnamefont{A.}~\bibnamefont{Pr\"ull}},
  \bibinfo{author}{\bibfnamefont{H.~G.} \bibnamefont{Evertz}},
  \bibnamefont{and} \bibinfo{author}{\bibfnamefont{W.}~\bibnamefont{von~der
  Linden}}, \bibinfo{journal}{Phys. Rev. B} \textbf{\bibinfo{volume}{66}},
  \bibinfo{pages}{144425} (\bibinfo{year}{2002}).

\bibitem[{\citenamefont{Yunoki and
  Moreo}(1998)}]{yunoki98:_static_dynam_proper_ferrom_kondo}
\bibinfo{author}{\bibfnamefont{S.}~\bibnamefont{Yunoki}} \bibnamefont{and}
  \bibinfo{author}{\bibfnamefont{A.}~\bibnamefont{Moreo}},
  \bibinfo{journal}{Phys. Rev. B} \textbf{\bibinfo{volume}{58}},
  \bibinfo{pages}{6403} (\bibinfo{year}{1998}).

\bibitem[{\citenamefont{Yunoki et~al.}(1998{\natexlab{a}})\citenamefont{Yunoki,
  Hu, Malvezzi, Moreo, Furukawa, and Dagotto}}]{yunoki98:_phase}
\bibinfo{author}{\bibfnamefont{S.}~\bibnamefont{Yunoki}},
  \bibinfo{author}{\bibfnamefont{J.}~\bibnamefont{Hu}},
  \bibinfo{author}{\bibfnamefont{A.~L.} \bibnamefont{Malvezzi}},
  \bibinfo{author}{\bibfnamefont{A.}~\bibnamefont{Moreo}},
  \bibinfo{author}{\bibfnamefont{N.}~\bibnamefont{Furukawa}}, \bibnamefont{and}
  \bibinfo{author}{\bibfnamefont{E.}~\bibnamefont{Dagotto}},
  \bibinfo{journal}{Phys. Rev. Lett.} \textbf{\bibinfo{volume}{80}},
  \bibinfo{pages}{845} (\bibinfo{year}{1998}{\natexlab{a}}).

\bibitem[{\citenamefont{Yi et~al.}(2000)\citenamefont{Yi, Hur, and
  Yu}}]{Yi_Hur_Yu:spinDE}
\bibinfo{author}{\bibfnamefont{H.}~\bibnamefont{Yi}},
  \bibinfo{author}{\bibfnamefont{N.~H.} \bibnamefont{Hur}}, \bibnamefont{and}
  \bibinfo{author}{\bibfnamefont{J.}~\bibnamefont{Yu}}, \bibinfo{journal}{Phys.
  Rev. B} \textbf{\bibinfo{volume}{61}}, \bibinfo{pages}{9501}
  (\bibinfo{year}{2000}).

\bibitem[{\citenamefont{Dagotto et~al.}(2001)\citenamefont{Dagotto, Hotta, and
  Moreo}}]{dagotto01:review}
\bibinfo{author}{\bibfnamefont{E.}~\bibnamefont{Dagotto}},
  \bibinfo{author}{\bibfnamefont{T.}~\bibnamefont{Hotta}}, \bibnamefont{and}
  \bibinfo{author}{\bibfnamefont{A.}~\bibnamefont{Moreo}},
  \bibinfo{journal}{Phys. Rep.} \textbf{\bibinfo{volume}{344}},
  \bibinfo{pages}{1} (\bibinfo{year}{2001}).

\bibitem[{\citenamefont{Motome and Furukawa}(2000)}]{Motome_Furukawa_3dDE}
\bibinfo{author}{\bibfnamefont{Y.}~\bibnamefont{Motome}} \bibnamefont{and}
  \bibinfo{author}{\bibfnamefont{N.}~\bibnamefont{Furukawa}},
  \bibinfo{journal}{J. Phys. Soc. Jpn.} \textbf{\bibinfo{volume}{69}},
  \bibinfo{pages}{3785} (\bibinfo{year}{2000}).

\bibitem[{\citenamefont{Motome and
  Furukawa}(2003{\natexlab{a}})}]{Motome_Furukawa_Ucl}
\bibinfo{author}{\bibfnamefont{Y.}~\bibnamefont{Motome}} \bibnamefont{and}
  \bibinfo{author}{\bibfnamefont{N.}~\bibnamefont{Furukawa}},
  \bibinfo{journal}{J. Phys. Soc. Jpn.} \textbf{\bibinfo{volume}{72}},
  \bibinfo{pages}{2126} (\bibinfo{year}{2003}{\natexlab{a}}).

\bibitem[{\citenamefont{Koller et~al.}(2003{\natexlab{a}})\citenamefont{Koller,
  Pr\"ull, Evertz, and von~der Linden}}]{KollerPruell2002b}
\bibinfo{author}{\bibfnamefont{W.}~\bibnamefont{Koller}},
  \bibinfo{author}{\bibfnamefont{A.}~\bibnamefont{Pr\"ull}},
  \bibinfo{author}{\bibfnamefont{H.~G.} \bibnamefont{Evertz}},
  \bibnamefont{and} \bibinfo{author}{\bibfnamefont{W.}~\bibnamefont{von~der
  Linden}}, \bibinfo{journal}{Phys. Rev. B} \textbf{\bibinfo{volume}{67}},
  \bibinfo{pages}{104432} (\bibinfo{year}{2003}{\natexlab{a}}).

\bibitem[{\citenamefont{Koller et~al.}(2003{\natexlab{b}})\citenamefont{Koller,
  Pr\"ull, Evertz, and von~der Linden}}]{KollerPruell2002c}
\bibinfo{author}{\bibfnamefont{W.}~\bibnamefont{Koller}},
  \bibinfo{author}{\bibfnamefont{A.}~\bibnamefont{Pr\"ull}},
  \bibinfo{author}{\bibfnamefont{H.~G.} \bibnamefont{Evertz}},
  \bibnamefont{and} \bibinfo{author}{\bibfnamefont{W.}~\bibnamefont{von~der
  Linden}}, \bibinfo{journal}{Phys. Rev. B} \textbf{\bibinfo{volume}{67}},
  \bibinfo{pages}{174418} (\bibinfo{year}{2003}{\natexlab{b}}).

\bibitem[{\citenamefont{Aliaga et~al.}(2001)\citenamefont{Aliaga, Normand,
  Hallberg, Avignon, and Alascio}}]{Aliaga_island_2d}
\bibinfo{author}{\bibfnamefont{H.}~\bibnamefont{Aliaga}},
  \bibinfo{author}{\bibfnamefont{B.}~\bibnamefont{Normand}},
  \bibinfo{author}{\bibfnamefont{K.}~\bibnamefont{Hallberg}},
  \bibinfo{author}{\bibfnamefont{M.}~\bibnamefont{Avignon}}, \bibnamefont{and}
  \bibinfo{author}{\bibfnamefont{B.}~\bibnamefont{Alascio}},
  \bibinfo{journal}{Phys. Rev. B} \textbf{\bibinfo{volume}{64}},
  \bibinfo{pages}{024422} (\bibinfo{year}{2001}).

\bibitem[{\citenamefont{Yunoki et~al.}(1998{\natexlab{b}})\citenamefont{Yunoki,
  Moreo, and Dagotto}}]{yunoki98:_phase_separ_induc_orbit_degrees}
\bibinfo{author}{\bibfnamefont{S.}~\bibnamefont{Yunoki}},
  \bibinfo{author}{\bibfnamefont{A.}~\bibnamefont{Moreo}}, \bibnamefont{and}
  \bibinfo{author}{\bibfnamefont{E.}~\bibnamefont{Dagotto}},
  \bibinfo{journal}{Phys. Rev. Lett.} \textbf{\bibinfo{volume}{81}},
  \bibinfo{pages}{5612} (\bibinfo{year}{1998}{\natexlab{b}}).

\bibitem[{\citenamefont{Hotta and Dagotto}(2000)}]{hotta00:_competition_fm_co}
\bibinfo{author}{\bibfnamefont{T.}~\bibnamefont{Hotta}} \bibnamefont{and}
  \bibinfo{author}{\bibfnamefont{E.}~\bibnamefont{Dagotto}},
  \bibinfo{journal}{Phys. Rev. B} \textbf{\bibinfo{volume}{61}},
  \bibinfo{pages}{11879} (\bibinfo{year}{2000}).

\bibitem[{\citenamefont{Hotta et~al.}(2001)\citenamefont{Hotta, Feiguin, and
  Dagotto}}]{hotta01:_stripes_oo_manganites}
\bibinfo{author}{\bibfnamefont{T.}~\bibnamefont{Hotta}},
  \bibinfo{author}{\bibfnamefont{A.}~\bibnamefont{Feiguin}}, \bibnamefont{and}
  \bibinfo{author}{\bibfnamefont{E.}~\bibnamefont{Dagotto}},
  \bibinfo{journal}{Phys. Rev. Lett.} \textbf{\bibinfo{volume}{68}},
  \bibinfo{pages}{4922} (\bibinfo{year}{2001}).

\bibitem[{\citenamefont{Motome and
  Furukawa}(2003{\natexlab{b}})}]{Motome_Furukawa_disorder}
\bibinfo{author}{\bibfnamefont{Y.}~\bibnamefont{Motome}} \bibnamefont{and}
  \bibinfo{author}{\bibfnamefont{N.}~\bibnamefont{Furukawa}},
  \bibinfo{journal}{Phys. Rev. Lett.} \textbf{\bibinfo{volume}{91}},
  \bibinfo{pages}{167204} (\bibinfo{year}{2003}{\natexlab{b}}).

\bibitem[{\citenamefont{Motome and
  Furukawa}(2003{\natexlab{c}})}]{Motome_Furukawa_disorder_b}
\bibinfo{author}{\bibfnamefont{Y.}~\bibnamefont{Motome}} \bibnamefont{and}
  \bibinfo{author}{\bibfnamefont{N.}~\bibnamefont{Furukawa}},
  \bibinfo{journal}{Phys. Rev. (} \textbf{\bibinfo{volume}{68}},
  \bibinfo{pages}{144432} (\bibinfo{year}{2003}{\natexlab{c}}).

\bibitem[{\citenamefont{Moreo et~al.}(1999)\citenamefont{Moreo, Yunoki, and
  Dagotto}}]{moreo_science_99}
\bibinfo{author}{\bibfnamefont{A.}~\bibnamefont{Moreo}},
  \bibinfo{author}{\bibfnamefont{S.}~\bibnamefont{Yunoki}}, \bibnamefont{and}
  \bibinfo{author}{\bibfnamefont{E.}~\bibnamefont{Dagotto}},
  \bibinfo{journal}{Science} \textbf{\bibinfo{volume}{283}},
  \bibinfo{pages}{2034} (\bibinfo{year}{1999}).

\bibitem[{\citenamefont{{M. Moraghebi} et~al.}(2001)\citenamefont{{M.
  Moraghebi}, {C. Buhler}, {S. Yunoki}, and {A. Moreo}}}]{Moraghebi_01:kondoCu}
\bibinfo{author}{\bibnamefont{{M. Moraghebi}}},
  \bibinfo{author}{\bibnamefont{{C. Buhler}}},
  \bibinfo{author}{\bibnamefont{{S. Yunoki}}}, \bibnamefont{and}
  \bibinfo{author}{\bibnamefont{{A. Moreo}}}, \bibinfo{journal}{Phys. Rev. B}
  \textbf{\bibinfo{volume}{63}}, \bibinfo{pages}{214513}
  (\bibinfo{year}{2001}).

\bibitem[{\citenamefont{{M. Moraghebi}
  et~al.}(2002{\natexlab{a}})\citenamefont{{M. Moraghebi}, {S. Yunoki}, and {A.
  Moreo}}}]{Moraghebi_02:kondoCu}
\bibinfo{author}{\bibnamefont{{M. Moraghebi}}},
  \bibinfo{author}{\bibnamefont{{S. Yunoki}}}, \bibnamefont{and}
  \bibinfo{author}{\bibnamefont{{A. Moreo}}}, \bibinfo{journal}{Phys. Rev. B}
  \textbf{\bibinfo{volume}{66}}, \bibinfo{pages}{214522}
  (\bibinfo{year}{2002}{\natexlab{a}}).

\bibitem[{\citenamefont{{M. Moraghebi}
  et~al.}(2002{\natexlab{b}})\citenamefont{{M. Moraghebi}, {A. Moreo}, and {S.
  Yunoki}}}]{Moraghebi_02b:kondoCu}
\bibinfo{author}{\bibnamefont{{M. Moraghebi}}},
  \bibinfo{author}{\bibnamefont{{A. Moreo}}}, \bibnamefont{and}
  \bibinfo{author}{\bibnamefont{{S. Yunoki}}}, \bibinfo{journal}{Phys. Rev.
  Lett.} \textbf{\bibinfo{volume}{88}}, \bibinfo{pages}{187001}
  (\bibinfo{year}{2002}{\natexlab{b}}).

\bibitem[{\citenamefont{Daghofer
  et~al.}(2004{\natexlab{a}})\citenamefont{Daghofer, Koller, Evertz, and
  von~der Linden}}]{DaghoferKoller2003}
\bibinfo{author}{\bibfnamefont{M.}~\bibnamefont{Daghofer}},
  \bibinfo{author}{\bibfnamefont{W.}~\bibnamefont{Koller}},
  \bibinfo{author}{\bibfnamefont{H.~G.} \bibnamefont{Evertz}},
  \bibnamefont{and} \bibinfo{author}{\bibfnamefont{W.}~\bibnamefont{von~der
  Linden}}, \bibinfo{journal}{J. Phys.: Condens. Matter}
  \textbf{\bibinfo{volume}{16}}, \bibinfo{pages}{1}
  (\bibinfo{year}{2004}{\natexlab{a}}).

\bibitem[{\citenamefont{Agterberg and Yunoki}(2000)}]{Agterberg_00}
\bibinfo{author}{\bibfnamefont{D.~F.} \bibnamefont{Agterberg}}
  \bibnamefont{and} \bibinfo{author}{\bibfnamefont{S.}~\bibnamefont{Yunoki}},
  \bibinfo{journal}{Phys. Rev. B} \textbf{\bibinfo{volume}{62}},
  \bibinfo{pages}{13816} (\bibinfo{year}{2000}).

\bibitem[{\citenamefont{Koshibae and Maekawa}(1998)}]{Yamanaka_98}
\bibinfo{author}{\bibfnamefont{M.~Y.~W.} \bibnamefont{Koshibae}}
  \bibnamefont{and} \bibinfo{author}{\bibfnamefont{S.}~\bibnamefont{Maekawa}},
  \bibinfo{journal}{Phys. Rev. Lett.} \textbf{\bibinfo{volume}{81}},
  \bibinfo{pages}{5604} (\bibinfo{year}{1998}).

\bibitem[{\citenamefont{Marinari}(1996)}]{mar:96}
\bibinfo{author}{\bibfnamefont{E.}~\bibnamefont{Marinari}},
  \emph{\bibinfo{title}{Optimized monte carlo methods}},
  \bibinfo{howpublished}{Lectures at the 1996 {B}udapest Summer School on Monte
  Carlo Methods} (\bibinfo{year}{1996}), \urlprefix\url{cond-mat/9612010}.

\bibitem[{\citenamefont{{K. Hukushima} and {K. Nemoto}}(1995)}]{huknem:95}
\bibinfo{author}{\bibnamefont{{K. Hukushima}}} \bibnamefont{and}
  \bibinfo{author}{\bibnamefont{{K. Nemoto}}},
  \bibinfo{journal}{cond-mat/9512035}  (\bibinfo{year}{1995}).

\bibitem[{\citenamefont{{K. Hukushima} et~al.}(1996)\citenamefont{{K.
  Hukushima}, {H. Takayama}, and {K. Nemoto}}}]{huknem:96}
\bibinfo{author}{\bibnamefont{{K. Hukushima}}},
  \bibinfo{author}{\bibnamefont{{H. Takayama}}}, \bibnamefont{and}
  \bibinfo{author}{\bibnamefont{{K. Nemoto}}}, \bibinfo{journal}{Int. J. Mod.
  Phys. C} \textbf{\bibinfo{volume}{7}}, \bibinfo{pages}{337}
  (\bibinfo{year}{1996}).

\bibitem[{\citenamefont{Chen and Allen}(2001)}]{allen01:sl_polaron}
\bibinfo{author}{\bibfnamefont{Y.-R.} \bibnamefont{Chen}} \bibnamefont{and}
  \bibinfo{author}{\bibfnamefont{P.~B.} \bibnamefont{Allen}},
  \bibinfo{journal}{Phys. Rev. B} \textbf{\bibinfo{volume}{64}},
  \bibinfo{pages}{064401} (\bibinfo{year}{2001}).

\bibitem[{\citenamefont{Dessau et~al.}(1998)\citenamefont{Dessau, Saitoh, Park,
  Shen, Villella, Hamada, Moritomo, and Tokura}}]{dessauI}
\bibinfo{author}{\bibfnamefont{D.~S.} \bibnamefont{Dessau}},
  \bibinfo{author}{\bibfnamefont{T.}~\bibnamefont{Saitoh}},
  \bibinfo{author}{\bibfnamefont{C.~H.} \bibnamefont{Park}},
  \bibinfo{author}{\bibfnamefont{Z.~X.} \bibnamefont{Shen}},
  \bibinfo{author}{\bibfnamefont{P.}~\bibnamefont{Villella}},
  \bibinfo{author}{\bibfnamefont{N.}~\bibnamefont{Hamada}},
  \bibinfo{author}{\bibfnamefont{Y.}~\bibnamefont{Moritomo}}, \bibnamefont{and}
  \bibinfo{author}{\bibfnamefont{Y.}~\bibnamefont{Tokura}},
  \bibinfo{journal}{Phys. Rev. Lett.} \textbf{\bibinfo{volume}{81}},
  \bibinfo{pages}{192} (\bibinfo{year}{1998}).

\bibitem[{\citenamefont{Saitoh et~al.}(2000)\citenamefont{Saitoh, Dessau,
  Moritomo, Kimura, Tokura, and Hamada}}]{dessauII}
\bibinfo{author}{\bibfnamefont{T.}~\bibnamefont{Saitoh}},
  \bibinfo{author}{\bibfnamefont{D.~S.} \bibnamefont{Dessau}},
  \bibinfo{author}{\bibfnamefont{Y.}~\bibnamefont{Moritomo}},
  \bibinfo{author}{\bibfnamefont{T.}~\bibnamefont{Kimura}},
  \bibinfo{author}{\bibfnamefont{Y.}~\bibnamefont{Tokura}}, \bibnamefont{and}
  \bibinfo{author}{\bibfnamefont{N.}~\bibnamefont{Hamada}},
  \bibinfo{journal}{Phys. Rev. B} \textbf{\bibinfo{volume}{62}},
  \bibinfo{pages}{1039} (\bibinfo{year}{2000}).

\bibitem[{\citenamefont{{Y.-D. Chuang} et~al.}(2001)\citenamefont{{Y.-D.
  Chuang}, {A. D Gromko}, {D. S. Dessau}, {T. Kimura}, and {Y.
  Tokura}}}]{dessauIII}
\bibinfo{author}{\bibnamefont{{Y.-D. Chuang}}},
  \bibinfo{author}{\bibnamefont{{A. D Gromko}}},
  \bibinfo{author}{\bibnamefont{{D. S. Dessau}}},
  \bibinfo{author}{\bibnamefont{{T. Kimura}}}, \bibnamefont{and}
  \bibinfo{author}{\bibnamefont{{Y. Tokura}}}, \bibinfo{journal}{Science}
  \textbf{\bibinfo{volume}{292}}, \bibinfo{pages}{1509} (\bibinfo{year}{2001}).

\bibitem[{\citenamefont{{J.H. Park} et~al.}(1996)\citenamefont{{J.H. Park}, {C.
  T. Chen}, {S-W. Cheong}, {W. Bao}, {G Meigs}, {V. Chakarian}, and {Y. U.
  Idzerda}}}]{Park95}
\bibinfo{author}{\bibnamefont{{J.H. Park}}}, \bibinfo{author}{\bibnamefont{{C.
  T. Chen}}}, \bibinfo{author}{\bibnamefont{{S-W. Cheong}}},
  \bibinfo{author}{\bibnamefont{{W. Bao}}}, \bibinfo{author}{\bibnamefont{{G
  Meigs}}}, \bibinfo{author}{\bibnamefont{{V. Chakarian}}}, \bibnamefont{and}
  \bibinfo{author}{\bibnamefont{{Y. U. Idzerda}}}, \bibinfo{journal}{Phys. Rev.
  Lett} \textbf{\bibinfo{volume}{76}}, \bibinfo{pages}{4215}
  (\bibinfo{year}{1996}).

\bibitem[{\citenamefont{Daghofer
  et~al.}(2004{\natexlab{b}})\citenamefont{Daghofer, Koller, von~der Linden,
  and Evertz}}]{DaghoferSCES}
\bibinfo{author}{\bibfnamefont{M.}~\bibnamefont{Daghofer}},
  \bibinfo{author}{\bibfnamefont{W.}~\bibnamefont{Koller}},
  \bibinfo{author}{\bibfnamefont{W.}~\bibnamefont{von~der Linden}},
  \bibnamefont{and} \bibinfo{author}{\bibfnamefont{H.~G.}
  \bibnamefont{Evertz}}, \bibinfo{journal}{cond-mat/0406635}
  (\bibinfo{year}{2004}{\natexlab{b}}).

\end{thebibliography}

\end{document}